\documentclass[usenatbib]{mnras}
\usepackage[active]{srcltx}
\usepackage{graphicx}
\usepackage{txfonts}
\usepackage{natbib}
\usepackage{multirow}
\usepackage{longtable}
\usepackage[english]{babel}
\usepackage[applemac]{inputenc}
\usepackage{rotating}
\usepackage{tablefootnote}
\usepackage{url}
\usepackage[multiple]{footmisc}
\usepackage{subfigure}

\usepackage{color}
\usepackage{soul}
\definecolor{jaune}{rgb}{1.0, 1.0, 0.0}

\definecolor{comment}{rgb}{1,0,0}
\definecolor{api}{rgb}{0,0.6,0.3}
\definecolor{oleg}{rgb}{0,0,1}
\definecolor{james}{rgb}{1,0.5,0.5}

\newcommand{\kms}{km\,s$^{-1}\,$}

\newcommand{\tn}{$\tau^{9}$\,Eri}

\def\gtrsim{\mathrel{\hbox{\rlap{\hbox{\lower4pt\hbox{$\sim$}}}\hbox{$>$}}}}
\def\ltsim{\mathrel{\hbox{\rlap{\hbox{\lower4pt\hbox{$\sim$}}}\hbox{$<$}}}}

\title[The pulsating magnetic SB2 \tn]{{\tn}: A bright pulsating magnetic Bp star in a 5.95-day double-lined spectroscopic binary}
\author[K. Woodcock et al.]{K. Woodcock,$^{1}$ G. A. Wade,$^1$ O. Kochukhov,$^2$ J. Sikora,$^{3}$ and A. Pigulski$^4$\\
$^{1}$Dept. of Physics and Space Science, Royal Military College of Canada, PO Box 17000 Station Forces, Kingston, ON, Canada K7K 0C6 \\
$^2$Department of Physics and Astronomy, Uppsala University, Box 516, SE-751 20 Uppsala, Sweden\\
$^3$Department of Physics and Astronomy, Bishop's University, Sherbrooke, Qu\'ebec J1M 1Z7, Canada\\
$^4$Astronomical Institute, University of Wroc{\l}aw, Kopernika 11, 51-622 Wroc{\l}aw, Poland
}
\date{Accepted XXX. Received YYY; in original form ZZZ}
\pubyear{2021}
\begin{document}
\label{firstpage}
\pagerange{\pageref{firstpage}--\pageref{lastpage}}
\maketitle

\begin{abstract}
\tn\ is a Bp star that was previously reported to be a single-lined spectroscopic binary. Using 17 ESPaDOnS spectropolarimetric (Stokes $V$) observations we identified the weak spectral lines of the secondary component and detected a strong magnetic field in the primary. We performed orbital analysis of the radial velocities of both components to find a slightly eccentric orbit ($e= 0.129$) with a period of $5.95382(2)$ days. 

The longitudinal magnetic field ($B_\ell$) of the primary was measured from each of the Stokes $V$ profiles, with typical error bars smaller than 10~G. Equivalent widths (EWs) of LSD profiles corresponding to only the Fe lines were also measured. We performed frequency analysis of both the $B_\ell$ and EW measurements, as well as of the Hipparcos, SMEI, and TESS photometric data. All sets of photometric observations produce two clear, strong candidates for the rotation period of the Bp star: 1.21 days and 3.82 days. The $B_\ell$ and EW measurements are consistent with only the 3.82-day period. We conclude that HD\,25267 consists of a late-type Bp star (M= $3.6_{-0.2}^{+0.1}~M_\odot$, T= $12580_{-120}^{+150}$~K) with a rotation period of 3.82262(4) days orbiting with a period of 5.95382(2) days with a late-A/early-F type secondary companion (M= $1.6\pm 0.1~M_\odot$, T= $7530_{-510}^{+580}$~K). The Bp star's magnetic field is approximately dipolar with $i= 41\pm 2\degr$, $\beta= 158\pm 5\degr$ and $B_{\rm d}= 1040\pm 50$~G. All evidence points to the strong $1.209912(3)$ day period detected in photometry, along with several other weaker photometric signals, as arising from $g$-mode pulsations in the primary.

\end{abstract}

\section{Introduction}

\tn\ (HD~25267) is a very bright ($V=4.66$ mag), nearby ($d=96$~pc), late-type Bp star that exhibits {significant} ($\sim$80~\kms) periodic radial velocity (RV) variations consistent with orbital motion in a binary system. The system has a long history of study. \citet{1908AN....177..171F} first documented HD~25267 as an RV variable. \citet{1927ApJ....65..300S} used 53 additional spectra to identify the system as an SB1 and found a period of 0.85437~d that did not seem to satisfy all of their data. Five additional RVs were published by \citet{1928PLicO..16....1C} that were employed by \citet{1928ApJ....67..399H} along with 10 new spectra to propose several alternative orbital periods, of which one (5.9542159~d) is close to the currently-accepted value. \citet{1950apJ...111..437S} used 38 radial velocity measurements to identify three possible orbital periods: 0.8542~d, 5.9542~d, and 1.1979~d, stating that the 5.9542~d period has `a slight advantage'. \citet{1999A&A...343..273L} combined Sahade's radial velocities with 6 of their own, reporting $5.9538 \pm 0.0001$~d as HD 25267's likely orbital period. 

\citet{1953MNRAS.113..357B} claimed detection of a magnetic field in this star (an effective magnetic field of $+1360\pm 700$~G), but later \citet{1958ApJS....3..141B} relegated \tn\ to his list of stars in which the presence of a magnetic field is probable but not firmly established. Additional magnetic measurements were reported by \citet{1980ApJS...42..421B}, ranging from $-350$~G to 0~G with typical uncertainties of 80-95~G, and proposed that the magnetic period was equal to the orbital period. 

Using 83 $uvby$ photometric data points taken in 1975 and 1977, \citet{1985A&A...144..251M} proposed two additional periods: 1.21 days (previously reported by \citealt{1981A&AS...46..151H}) and 3.8 days. These results were confirmed by \citet{1991A&A...248..179C}. Most recently, \citet{2020MNRAS.493.3293B} identified the 1.21~d period as the rotational period, while \citet{2017A&A...601A..14M} reasserted that the 5.95~d period represents both the rotation and orbit periods (in agreement with {\citealt{1980ApJS...42..421B}}), and that the origins of the 1.21~d and 3.82~d periods remain unclear.

Three periods (1.21~d, 3.82~d, and 5.95~d days) {consistently} recur in modern analyses of the system. The 5.95~d period seems firmly established to represent the orbital period. It may also represent the rotational period of the Bp star (as proposed by \citealt{1980ApJS...42..421B} and \citealt{2017A&A...601A..14M}). On the other hand, the prominent (in photometry) 1.21~d or 3.82~d periods might be the star's rotational period. In this paper we analyze high resolution spectropolarimetric observations of HD~25267 to detect the spectral lines of the secondary star, revealing the system to be a double-lined spectroscopic binary (SB2). We determine the radial velocities of both components to model the 5.95~d orbit of the system and constrain the physical parameters of both stars. We exploit measurements of the longitudinal magnetic field ($B_\ell$) and equivalent widths of the primary's mean spectral line to establish its 3.82~d rotational period and to model its magnetic field geometry. Finally, we evaluate possible origins of the 1.21~d period.

\section{Observations}

\subsection{Spectropolarimetry}

Our investigation takes advantage of 17 spectropolarimetric (Stokes $V$) observations taken using ESPaDOnS at the Canada-France-Hawaii Telescope (CFHT). Four observations, acquired from the CFHT archive, were obtained on August 20, September 23, 24, and 25, 2013. Thirteen new observations took place almost nightly from December 26, 2017 to January 9, 2018. The spectra, reduced at the CFHT with the Upena pipeline feeding the Libre-Esprit reduction software, span a wavelength range from 369\,--\,1048~nm with a resolving power of 65,000, and a median signal-to-noise ratio (S/N) of approximately 850 per pixel. The spectra were normalized to the continuum by performing iterative fitting of continuum points in the unmerged orders using polynomial fits. Further details about the acquisition and reduction of the data are provided by, e.g., \citet{2016MNRAS.456....2W}. The log of spectropolarimetric observations is reported in Table~\ref{tab:Log}.

\begin{table*}
\caption{\label{tab:Log}Log of spectropolarimetric observations, including orbital and rotational phases (columns 3 and 4), longitudinal magnetic field measured from Stokes $V$ and diagnostic null (columns 5 and 6), equivalent widths (from Fe LSD profiles, column 7), exposure time and signal-to-noise ratio (columns 8 and 9), and measured radial velocities (columns 10 and 11). Orbital and rotational phases are computed using the ephmerides described in Eqs. (2) and (3).} 
\begin{tabular}{ccccrrccrrr}
\hline
Spectrum & HJD & $\phi_{\rm orb}$ & $\phi_{\rm rot}$ & $B_{\ell} (V)$ & $B_{\ell} (N)$ & EW  & Exp.& S/N & RV$_{\rm pri}$  & RV$_{\rm sec}$ \\
ID &  &  &  & (G) & (G) & ({\kms}) & (s) & (pix$^{-1}$) &  \multicolumn{2}{c}{(km\ ${\rm s}^{-1}$)}\\
\hline
(1) & (2) & (3) & (4) & (5) & (6) & (7) & (8) & (9) & (10) & (11)\\
\hline
1648693 & 2456525.142 & 0.646 & 0.061 & $-$142 $\pm$ 8 & $-$2 $\pm$ 6 & 1.525 $\pm$ 0.004 & 110 & 929 & 44 & $-$29\\
1655882 & 2456559.139 & 0.356 & 0.955 & $-$138 $\pm$ 8 & 4 $\pm$ 8 & 1.484 $\pm$ 0.004 & 110 & 808 & 48 & $-$42\\
1656054 & 2456560.009 & 0.502 & 0.183 & $-$214 $\pm$ 9 & 11 $\pm$ 9 & 1.590 $\pm$ 0.007 & 110 & 758 & 57 & $-$59\\
1656328 & 2456560.984 & 0.666 & 0.438 & $-$261 $\pm$ 7 & 3 $\pm$ 7 & 1.631 $\pm$ 0.004 & 110 & 866 & 41 & $-$24\\
2237404 & 2458114.712 & 0.629 & 0.895 & $-$145 $\pm$ 18 & 4 $\pm$ 18 & 1.459 $\pm$ 0.003 & 180 & 372 & 46 & $-$34\\
2237408 & 2458114.722 & 0.630 & 0.898 & $-$162 $\pm$ 10 & $-$5 $\pm$ 9 & 1.465 $\pm$ 0.003 & 180 & 747 & 46 & $-$34\\
2237413 & 2458114.740 & 0.634 & 0.902 & $-$148 $\pm$ 13 & 13 $\pm$ 13 & 1.471 $\pm$ 0.003 & 180 & 555 & 45 & $-$33\\
2237541 & 2458115.787 & 0.809 & 0.176 & $-$198 $\pm$ 9 & 0 $\pm$ 9 & 1.479 $\pm$ 0.003 & 180 & 767 & 8 & 45\\
2237725 & 2458116.731 & 0.968 & 0.423 & $-$238 $\pm$ 10 & 13 $\pm$ 10 & 1.560 $\pm$ 0.004 & 180 & 773 & $-$22 & 120\\
2237881 & 2458117.747 & 0.139 & 0.689 & $-$231 $\pm$ 7 & $-$3 $\pm$ 6 & 1.528 $\pm$ 0.003 & 180 & 1080 & 0 & 71\\
2238075 & 2458118.770 & 0.310 & 0.957 & $-$137 $\pm$ 7 & 3 $\pm$ 7 & 1.559 $\pm$ 0.003 & 180 & 894 & 41 & $-$24\\
2238187 & 2458119.762 & 0.477 & 0.216 & $-$207 $\pm$ 6 & 4 $\pm$ 5 & 1.564 $\pm$ 0.003 & 180 & 1137 & 56 & $-$59\\
2238459 & 2458120.827 & 0.656 & 0.495 & $-$260 $\pm$ 6 & 0 $\pm$ 6 & 1.613 $\pm$ 0.003 & 180 & 917 & 49 & $-$25\\
2238647 & 2458121.788 & 0.817 & 0.746 & $-$230 $\pm$ 6 & $-$2 $\pm$ 6 & 1.467 $\pm$ 0.003 & 180 & 975 & 7 & 51\\
2238771 & 2458122.793 & 0.986 & 0.009 & $-$131 $\pm$ 11 & $-$2 $\pm$ 10 & 1.509 $\pm$ 0.003 & 180 & 549 & $-$24 & 121\\
2238898 & 2458123.829 & 0.160 & 0.280 & $-$227 $\pm$ 5 & $-$4 $\pm$ 5 & 1.577 $\pm$ 0.003 & 180 & 1080 & 4 & 57\\
2239595 & 2458128.830 & 0.000 & 0.588 & $-$281 $\pm$ 7 & $-$6 $\pm$ 6 & 1.544 $\pm$ 0.004 & 180 & 1089 & $-$24 & 123\\
\hline
\end{tabular}
\end{table*}

\begin{table}
\caption{\label{tab:periods}A summary of most significant periods (in days) detected in the various magnetic, spectroscopic, and photometric datasets.}
\centering
\begin{tabular}{clll}
\hline
Dataset & \multicolumn{1}{c}{1.21} & \multicolumn{1}{c}{3.82} & \multicolumn{1}{c}{5.95} \\
\hline
Hipparcos & 1.2100(4) & 3.8227(11) & \multicolumn{1}{c}{---} \\
TESS & 1.209921(15) & 3.81879(15) & 5.928(4)\\
SMEI  & 1.209912(3) & 3.82262(4) & \multicolumn{1}{c}{---}\\
$B_\ell$  & \multicolumn{1}{c}{---} & 3.8230(6) & \multicolumn{1}{c}{---} \\
EW  & \multicolumn{1}{c}{---} & 3.8230(5) & \multicolumn{1}{c}{---} \\
RV  & \multicolumn{1}{c}{---} & \multicolumn{1}{c}{---} & 5.95382(2) \\
\hline
\end{tabular}
\end{table}

\begin{table*}
\caption{\label{tab:periods2}Parameters of the least-squares frequency fits to the SMEI and TESS photometric datasets. Phases are given in radians, for the following epochs: HJD 2454000.0 for SMEI and BJD 2458440.0 for TESS. The signal-to-noise ratio (SNR) was calculated by dividing amplitude (S) by the noise (N) calculated locally using frequency spectra of the residuals from the presented multi-frequency fits. ``Notes" describe the proposed origin and relationship to any other detected signals.}
\centering
\begin{tabular}{clclrlclrc}
\hline
& \multicolumn{4}{c}{SMEI} & \multicolumn{4}{c}{TESS} & \\
\hline
ID & \multicolumn{1}{c}{Frequency} & Amplitude & \multicolumn{1}{c}{Phase} & \multicolumn{1}{c}{SNR} & \multicolumn{1}{c}{Frequency} & Amplitude & \multicolumn{1}{c}{Phase} & \multicolumn{1}{c}{SNR} & Notes\\
& \multicolumn{1}{c}{(d$^{-1}$)} & (mmag) & \multicolumn{1}{c}{(rad)} && \multicolumn{1}{c}{(d$^{-1}$)} & (mmag) & \multicolumn{1}{c}{(rad)} &&\\
\hline
$f_1$ & 0.8265067(21) & 8.11(9) & 0.745(11) & 74.5 & 0.826500(10) & 8.938(9) & 5.0816(10) & 70.4 & $g$-mode \\
$f_2$ & 0.2616008(24) & 6.88(9) & 5.166(13) & 49.8 & 0.261863(10) & 8.379(9) & 2.0744(10) & 49.0 & $f_{\rm rot}$ \\
$f_3$ & 0.5232017 & 1.08(9) & 4.75(8) & 8.6 & 0.523727(2) & 1.726(9) & 4.920(5) & 12.8 & $2\,f_{\rm rot}$ \\
$f_4$ & 0.879600(18) & 0.99(9) & 2.28(9) & 9.4 & 0.88090(8) & 1.224(8) & 5.000(7) & 9.7 &  $g$-mode\\
$f_5$ & 0.257385(23) & 0.75(9) & 5.96(11) & 5.4 & \multicolumn{1}{c}{---} & --- & \multicolumn{1}{c}{---} & \multicolumn{1}{c}{---} & spurious\\
$f_6$ & 0.564906 & 0.56(9) & 3.63(15) & 4.5 & 0.564636(2) & 0.468(9) & 5.152(18) & 3.5 & $f_1-f_{\rm rot}$ \\ 
$f_7$ & 0.61531(4) & 0.55(9) & 1.63(16) & 4.5 & 0.61498(16) & 0.553(8) & 2.917(15) & 4.8 & $g$-mode\\
$f_8$ & \multicolumn{1}{c}{---} & --- & \multicolumn{1}{c}{---} & --- & 1.01773(11) & 0.852(8) & 2.818(10) & 7.8 & $g$-mode \\
$f_9$ & \multicolumn{1}{c}{---} & \multicolumn{1}{c}{---} & \multicolumn{1}{c}{---} & --- & 1.10199(16) & 0.563(9) & 0.543(15) & 4.8 &  $g$-mode\\
$f_{10}$ & \multicolumn{1}{c}{---}& ---& \multicolumn{1}{c}{---}& --- & 0.16868(11) & 0.842(9) & 5.531(10) & 4.8 & $f_{\rm orb}$\\
\hline
\end{tabular}
\label{tab:my_label}
\end{table*}

\subsection{Photometry}

Our investigation employed Hipparcos photometry of HD~25267 extracted from the Centre de Donn\'ees Astronomiques de Strasbourg (CDS). A total of 193 data points were observed over more than 3 years, and 2 outliers were removed from the analysis. The typical precision was of a few mmag.

The Solar Mass Ejection Imager (SMEI) experiment \citep{2003SoPh..217..319E,2004SoPh..225..177J} was placed on-board the Coriolis spacecraft and was aimed at measuring sunlight scattered by free electrons in the solar wind. We used photometry of HD~25267 obtained during nearly 8 years and available through the University of California San Diego (UCSD) web page\footnote{http://smei.ucsd.edu/new\_smei/index.html}. The SMEI time series are affected by long-term calibration effects, especially a repeatable variability with a period of one year. {The raw SMEI UCSD photometry of HD~25267 was corrected for the one-year variability by subtracting an interpolated mean light curve, which was obtained by folding the raw data with the period of one year, calculating median values in 200 intervals in phase, and then interpolating between them. In addition, the worst parts of the light curve and outliers were removed. The data points were also assigned individual uncertainties (typically 5-10 mmag) calculated using the scatter of the neighbouring data. Then, a model consisting of the dominant frequencies was fit to the data. Finally, the low-frequency instrumental variability was filtered out by subtracting a trend using residuals from the fit. The last two steps were iterated several times, yielding 22771 data points that were ultimately analyzed.} 

Finally, we employed new photometry from the Transiting Exoplanet Survey Satellite (TESS). The primary goal of the NASA's TESS mission \citep{2014SPIE.9143E..20R,2015JATIS...1a4003R} is the detection of planets by means of the transit method. TESS observations cover almost the entire sky, excluding only the regions with low Galactic latitudes ($|b|<$~6$\degr$). Observations were carried out with a 30-min cadence, but selected stars, including HD~25267, were observed with a shorter, 2-min cadence. The star was observed in Sectors 4 and 5. The observations spanned {54}~d between {October 19, 2018} and {December 11, 2018}, and consisted of {33662} data points. In the subsequent analysis we used SAP fluxes and removed all data points with quality flag different from 0.

\section{Least-Squares Deconvolution} \label{LSD Profiles}

We employed Least-Squares Deconvolution \citep[LSD;][]{1997MNRAS.291..658D,2010A&A...524A...5K} to compute high signal-to-noise ratio (S/N) pseudo-line profiles following the essential procedure laid out by \citet{2017MNRAS.465.2432G}. LSD was applied to each spectrum to compute mean line profiles with a high S/N. A line mask is produced using `Extract Stellar' requests from the Vienna Atomic Line Database \citep{1995A&AS..112..525P}, resulting in a list of all absorption lines and related data expected to be present in a star at a given temperature and surface gravity. The line mask was cleaned and tweaked using an interactive graphical tool developed by Jason Grunhut (see e.g. \citealt{2017MNRAS.465.2432G}).

\begin{figure}
\centering
\includegraphics[width=\columnwidth]{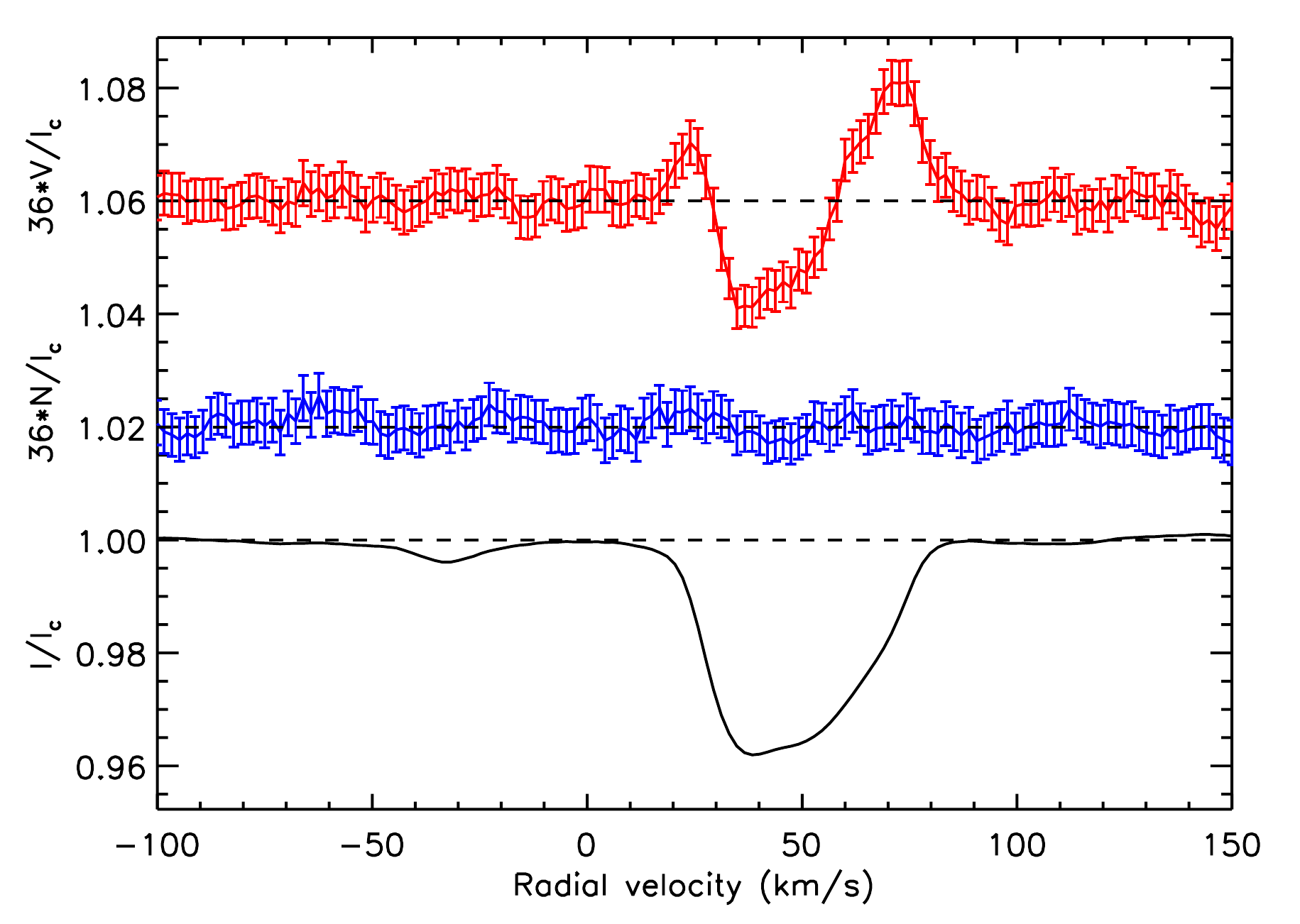}
\caption{Stokes $V$ (red), diagnostic null $N$ (blue), and Stokes $I$ (black) LSD profiles extracted from one of the spectropolarimetric observations of HD~25267. The secondary's line profile is visible at $-35$~{\kms}. The Stokes $V$ and $N$ profiles have been scaled and shifted for display purposes.}
\label{fig:2238187LSD}
\end{figure}

Line masks of various effective temperatures --- ranging from 7 to 14~kK --- were tested on the spectrum. We were able to identify two spectral line features in our LSD profiles; a strong line produced by the Bp primary star, and a second, weaker line, corresponding to the secondary component of the system. The 10\,kK mask was selected for our analysis, as it produced the deepest secondary Stokes $I$ profile. All LSD Stokes $V$ profiles (see e.g.~Fig.~\ref{fig:2238187LSD}) show clear Zeeman signatures in the mean spectral line of the primary star, indicative of the presence of a magnetic field.

In addition to LSD profiles extracted using the full mask, we also extracted LSD profile sets using a sub-mask restricted to the lines of Fe.

\section{Measurements}

All measurements are reported in Table~\ref{tab:Log}.

\subsection{Radial velocities}\label{sect:vr}

Radial velocities of the primary and secondary stars were determined by fitting a Gaussian function to both lines in each Stokes $I$ LSD profile. Typical uncertainties (inferred from the RMS scatter about the best-fit orbital model derived in Sect.~\ref{Sect:orbit}) are about 1\,\kms. We were able to successfully measure radial velocities for both components in all 17 observations.

\subsection{Longitudinal magnetic field}

To calculate the mean longitudinal magnetic field of the primary star, we measured the first-order moment of the Stokes $V$ profile normalized to the equivalent width of the Stokes $I$ profile according to the expression:
\begin{equation}
B_\ell = -2.14\times10^{11} \frac{\int(v-v_0)V(v)\mbox{d}v}{\lambda zc \int [1-I(v)]\mbox{d}v}
\end{equation}
\citep{1997MNRAS.291..658D,2000MNRAS.313..851W}, where $\lambda$ represents the wavelength in nm, $z$ is the Land\'e factor, $c$ is the speed of light, and $V(v)$ and $I(v)$ are the Stokes $V$ and $I$ profile intensities, respectively. $v_0$ is the radial velocity of the  center of gravity of the Stokes $I$ profile. The equation was evaluated in a range $\pm 40$\,\kms about the primary's center-of-gravity. 

\subsection{Equivalent widths}

The Stokes $I$ profiles of the primary star exhibit significant variability. We measured the equivalent widths (EWs) of the Fe Stokes $I$ LSD profiles by first centering each profile of the primary star at $v=0$\,\kms, then performing local renormalization of the continuum. We then performed trapezoidal integration of the intensity between continuum and the profile in the velocity range $\pm 40$\,\kms. 

\section{Period analysis}\label{Sect:periods}

We performed period analysis of the RV, photometric, $B_\ell$ and EW measurements. We used a Lomb-Scargle approach for the RVs, EWs, and $B_\ell$ data, while we employed the Fourier approach of {\sc Period04} for the photometric data. The key periods recovered from each data set are summarized in Table \ref{tab:periods}. Details of frequencies detected in the SMEI and TESS timeseries are reported in Table~\ref{tab:periods2}.

All three photometric datasets display 2 similarly-strong peaks in their periodograms near 0.826\,d$^{-1}$ and 0.262\,d$^{-1}$, corresponding to periods of 1.21 and 3.82 d, respectively. 

The Hipparcos photometry shows significant peaks yielding frequencies of $f_1=0.8264(3)$\,d$^{-1}$ and $f_2=0.26160(7)$\,d$^{-1}$ (corresponding to periods of $1.2100(4)$ days and $3.822(1)$ days, with respective amplitudes of 10.8 and 8.5 mmag), consistent with the results of \citet{1985A&A...144..251M}. 

\begin{figure}
\centering
\hspace{-0.25cm}\includegraphics[width=0.95\columnwidth]{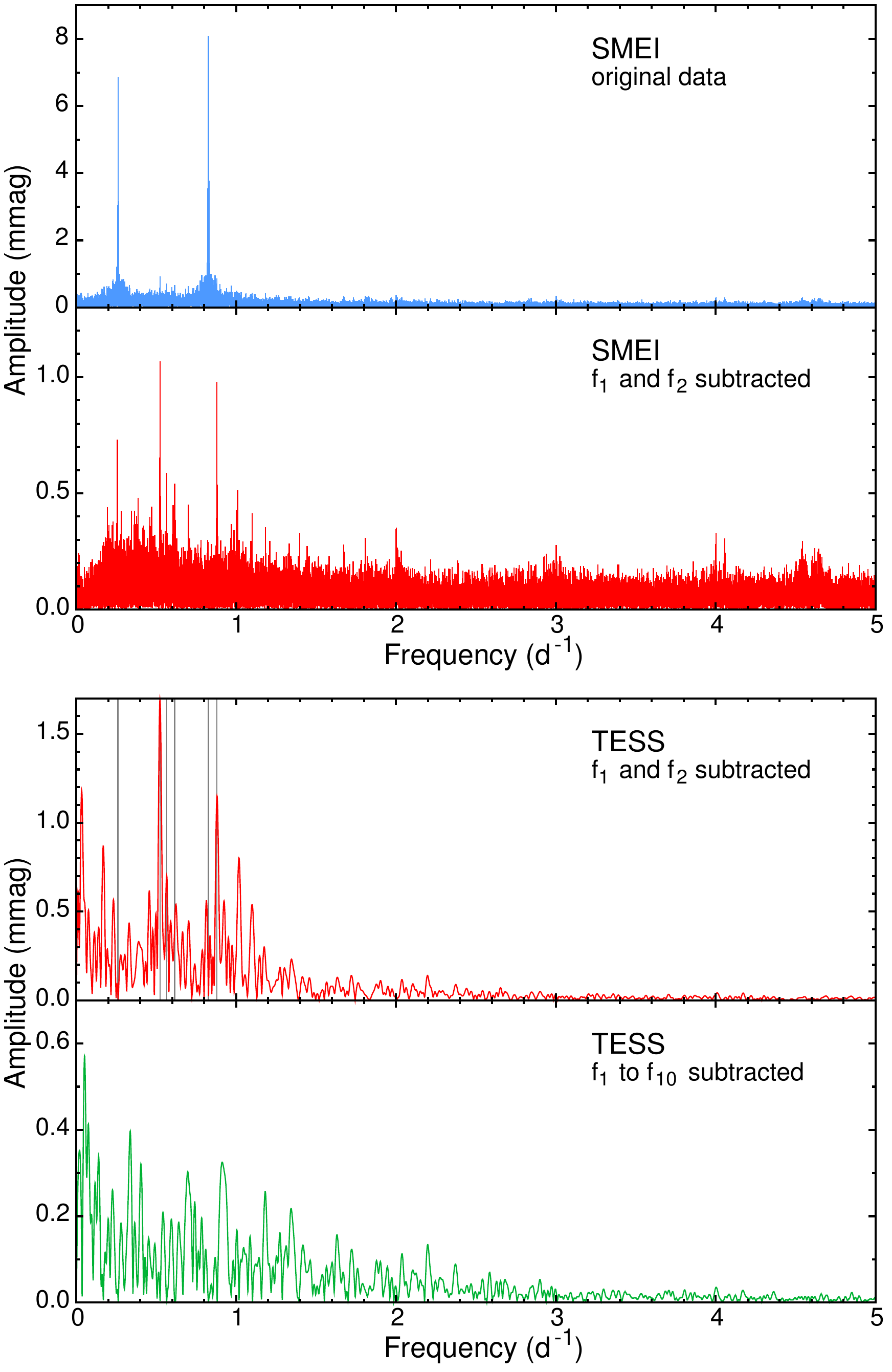}
\caption{Frequency spectra of the photometric data of \tn. {\it Top panels:} Frequency spectra of the original SMEI data (blue) and after subtraction of the two dominant frequencies, $f_1$ and $f_2$ (red). {\it Bottom panels:} Frequency spectra of TESS data after subtraction of $f_1$ and $f_2$ (red) and after subtracting all terms listed in Table~\ref{tab:periods2} (green). Grey vertical lines in the upper panel mark six frequencies (excluding $f_5$, deemed to be instrumental) detected in the SMEI data and listed in Table \ref{tab:periods2}.}
\label{fig:photperiodograms}
\end{figure}

A significant advantage of the SMEI photometry is excellent frequency resolution. The data, however, suffer from some instrumental effects. Again, the amplitude spectrum of the data is dominated by two frequencies, $f_1 = 0.8265067(21)$\,d$^{-1}$ and $f_2 = 0.2616008(24)$\,d$^{-1}$, corresponding to periods of 1.209912(3)~d (amplitude of 8.11(9) mmag) and 3.82262(4)~d (amplitude of 6.88(9) mmag). The amplitude spectrum after prewhitening these two dominating terms is also interesting; see the upper panel of Fig.~\ref{fig:photperiodograms}. There are at least two peaks that are definitely significant (Table \ref{tab:my_label}). The first, at $f_3 = 0.5232017$\,d$^{-1}$ (amplitude 1.08(9) mmag), is the harmonic of $f_2$ ($f_3 = 2\,f_2$). The other, $f_4 = 0.879600$\,d$^{-1}$ (amplitude of 0.99(9) mmag), is independent. There is no doubt that $f_3$ is an harmonic of $f_2$ given the superb resolution of the SMEI data.

The SMEI amplitude spectrum also shows weaker signals. They are of lower significance, but generally likely significant: $f_5 = 0.257385$\,d$^{-1}$ (0.75(9) mmag) could be a $g$-mode based on the magnitude of its frequency, but is more likely a remnant of the detrending procedure. The noise level in SMEI data increases towards low frequencies, hence we performed detrending which reduces signals at the lowest frequencies (the drop of signal below 0.25\,d$^{-1}$). Therefore, we doubt that $f_5$ is real. $f_6 = 0.564906$\,d$^{-1}$ (amp. 0.56(9) mmag) is a combination frequency ($f_6 = f_1 - f_2$). It is still strong enough to be significant, but more importantly, its frequency is exactly the difference between the two dominant frequencies. This provides important insights into the origins of $f_1$ and $f_2$ that will be discussed later in this paper.  Finally, we identify $f_7 = 0.61531$\,d$^{-1}$ (0.55(9) mmag), which is again possibly a $g$-mode based on the magnitude of its frequency. Some other smaller peaks are also present in the spectrum, but given that many are close to $1, 2, 3$, and $4$\,d$^{-1}$ as well as the cluster of peaks at $\sim$4.6\,d$^{-1}$ --- where such clusters at frequencies between 3\,--\,5\,d$^{-1}$ show up in almost all SMEI data --- we conclude that they are likely spurious.

The TESS data are much more precise than SMEI, but also suffer from some instrumental effects. Two fluxes are available in photometry, the {\sc SAP$\_$FLUX} (hereafter SAP) and the corrected {\sc PDCSAP$\_$FLUX} (hereafter PDC). There is a large drop in flux in SAP after the mid-time gap in the Sector 4 observations. It is largely corrected in the PDC flux, but not perfectly. We decided to use the SAP flux, because from our experience analyzing TESS data we conclude that the PDC correction sometimes introduces unwanted signals. We only removed a small part of the timeseries between TJD 1420 and 1424 (where $\mbox{TJD}=\mbox{BJD}-2457000$).

The spectrum of the TESS data is dominated by the same two frequencies as the SMEI data, $f_1$ and $f_2$. Given the results from SMEI, we were interested in the signals which remain after subtraction of these two terms. The residual frequency spectrum of the TESS data, after prewhitening these two terms, is shown in Fig.~\ref{fig:photperiodograms}. We identify the following significant frequencies: $f_3$ (at 0.523727\,d$^{-1}$ $=2\,f_2$) is clearly present; $f_4$ (0.88090\,d$^{-1}$, a likely $g$-mode) is present as well. $f_6$ (0.564636\,d$^{-1}$ = $f_2-f_1$) and $f_7$ (0.61498\,d$^{-1}$) are barely seen because of the poor frequency resolution, but are possibly present. $f_5$ (0.257385\,d$^{-1}$) is absent, which confirms our suspicion that it is an artefact in the SMEI data. At least two more peaks are present, $f_8$ at 1.01773\,d$^{-1}$ and $f_9$ at 1.10199\,d$^{-1}$. Both can be seen in the SMEI data, although the frequencies close to 1\,d$^{-1}$ are affected by instrumental effects in the SMEI data.

There are also two peaks at low frequencies in the TESS data. The first one, at 0.033\,d$^{-1}$ (not included in Table \ref{tab:my_label}), is likely of instrumental origin because there is a small jump between the data before and after the gap. The other one, $f_{10}$, is located at 0.16868\,d$^{-1}$ and corresponds to the known RV period. Hence, the photometry appears to be weakly modulated according to the 5.95-day orbital period; this will be discussed in the next Section. 

It is clear that the residuals in the TESS photometry after subtracting the above nine terms (Fig.\,\ref{fig:photperiodograms}) still include significant variability, which can be seen both in the residual light curve and its frequency spectrum. This will be discussed later in the paper.

We next combined our RV measurements with 44 published velocities of the primary component reported by \citet{1950apJ...111..437S} (38 measurements) and \citet{1999A&A...343..273L}  (6 measurements). Performing period analysis independently on the primary and secondary RVs yields similar results. The lowest reduced $\chi^2$ in the primary's periodogram is located at $5.95382(2)$~d, while the secondary's periodogram gives $5.9538(4)$ days. 

The periodogram of our $B_\ell$ measurements exhibits many close peaks forming a broad envelope centered near $3.8$ days. The peak most compatible with the frequencies detected in the photometric data is located at $3.8230(6)$ days. Additional power is present near 1.4~d, the daily alias of 3.8~d. $B_\ell$ measurements by \citet{1980ApJS...42..421B} suggested rotational periods near 0.85 and 5.95 days. Our $B_\ell$ data rule out periods near both of these values, as well as periods near 1.21~d. This is discussed further in Sect.~\ref{sect:rotandmag}.

Finally, we analyzed the EWs measured from the Fe LSD profiles. As with the $B_\ell$ data, the periodogram exhibits a broad envelope composed of sharp peaks centered near 3.8~d, again with some power near 1.4~d. The peak most compatible with the photometric and $B_\ell$ periods is located at 3.823(1) days.

\section{Modeling of the orbit}
\label{Sect:orbit}

We used the IDL orbital fitting code {\sc Xorbit} to model the orbit. This code determines the best-fitting orbital period $P_{\rm orb}$, time of periastron passage ($T_0$), eccentricity ($e$), longitude of the periastron ($\omega$), semi-amplitudes of each component's radial velocities ($K_1$ and $K_2$), and the radial velocity of the center of mass ($\gamma$) \citep{1992ASPC...32..573T}, performing least-squares fits to the measured radial velocities. Using the RV period identified in Sect.~\ref{Sect:periods} as an initial guess for $P_{\rm orb}$, and using the range of measured RVs as initial guesses for $K_{\rm 1}$ and $K_{\rm 2}$, the solution converged quickly to an acceptable, stable fit.

\begin{table}
\caption{Final orbital solution for $\tau^9$~Eri: results of simultaneous modeling of the RV variations of the primary and secondary components.}\label{orbital_params}
\centering
\begin{tabular}{lrr}
\hline
Quantity & Value & Uncertainty \\
\hline
$P_{\rm orb}$ (d) & 5.95382 & 0.00002 \\
$T_0$ (HJD) & 2456991.65 & 0.08\\
$K_1$ ({\kms}) & 40.0 & 0.6\\
$K_2$ ({\kms}) & 89.9 & 1.4\\
$\gamma$ ({\kms}) & 21.1 & 0.4\\
$e$ & 0.129 & 0.010\\
$\omega$ ($\degr$) & 183.2 & 4.3\\
RMS$_1$ ({\kms}) & 0.7\\
RMS$_2$ ({\kms}) & 1.0 \\
$M_1/M_2$ & 2.25& 0.07 \\
$M_1\sin^3 i$ ($M_\odot$) & 0.91& 0.05 \\
$M_2\sin^3 i$ ($M_\odot$)& 0.41 & 0.03 \\
$a\sin i$ (AU) & 0.0705  & 0.0012\\
\hline
\end{tabular}
\end{table}

\begin{figure}
\centering
\hspace{-0.4cm}\includegraphics[width=1.02\columnwidth]{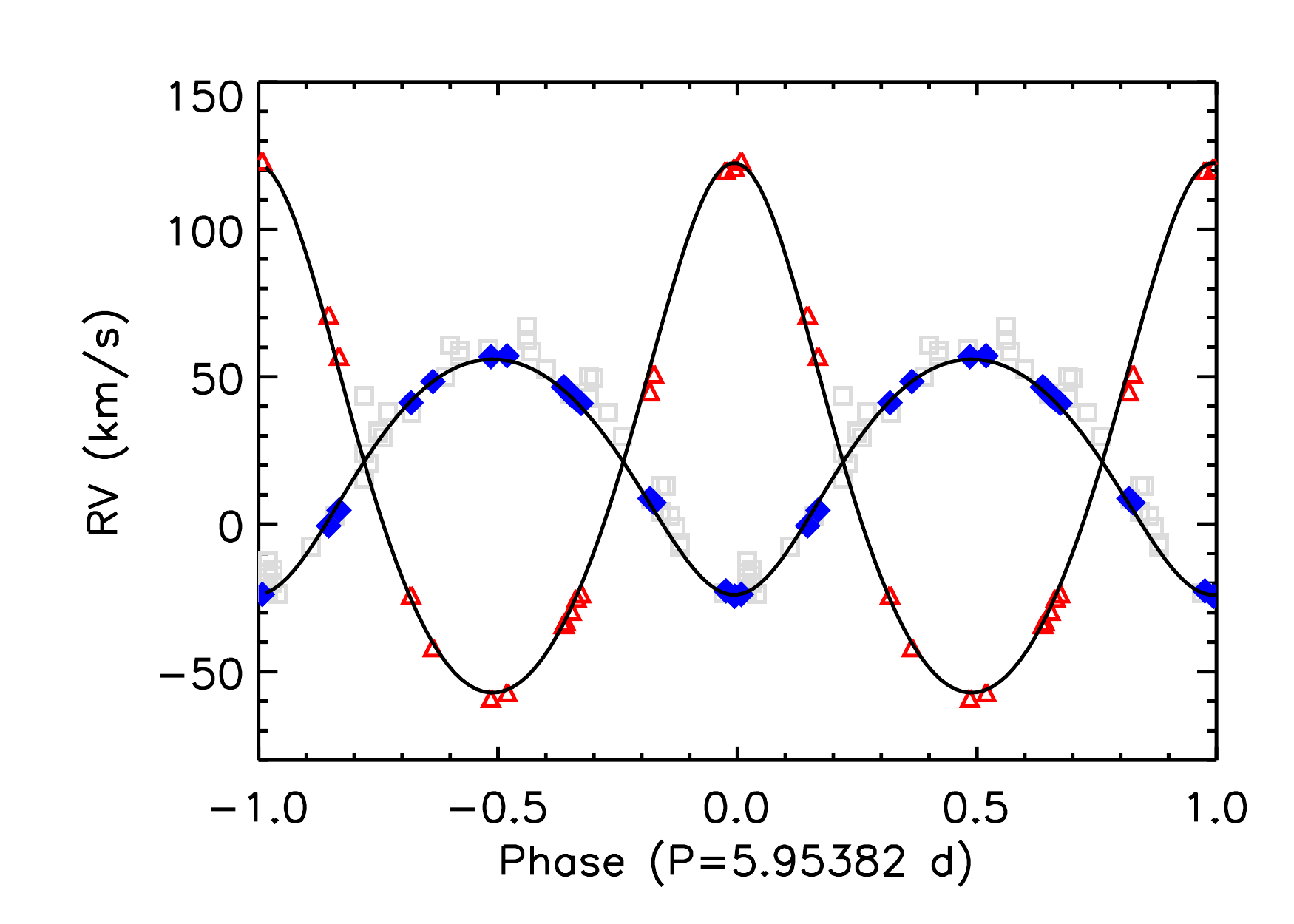}
\caption{Radial velocity and theoretical velocity curves of the primary (blue) and secondary (red). Grey squares indicate published velocities of the primary. The ephemeris used for phasing is given by Eq.\,(\ref{eqorb}).}
\label{fig:Orbital Modeling}
\end{figure}

Figure~\ref{fig:Orbital Modeling} illustrates the final model fit to the phased RV measurements according to the orbital ephemeris:
\begin{equation}\label{eqorb}
T_{0}(E_{\rm orb}) = \mbox{HJD}\, 2456991.65(8) + 5.95382(2)\cdot {E_{\rm orb}},
\end{equation}
where $T_0(E_{\rm orb})$ represents the epoch of periastron passage and $E_{\rm orb}$ is the number of orbital cycles elapsed from $T_0$. 

The measurements of both the primary and secondary components exhibit coherent variations with an RMS scatter of approximately 1 {\kms} about the model. The RVs imply a small but significant eccentricity $e=0.129 \pm 0.010$. The semi-amplitudes $K_{\rm 1}=40.0\pm 0.6$~{\kms} and $K_{\rm 2}=89.9\pm 1.4$~{\kms} imply a mass ratio $M_{\rm 1}/M_{\rm 2}=2.25\pm 0.07$. The total system mass is $(M_{\rm 1}+M_{\rm 2})=(1.32\pm 0.06)\,\sin^{-3} i~M_\odot$, with individual masses of $M_{\rm 1}=(0.91\pm 0.05)\,\sin^{-3} i~M_\odot$ and $M_{\rm 2}=(0.41\pm 0.03)\,\sin^{-3} i~M_\odot$. The projected semi-major axis is $a\sin i=0.0705\pm 0.0012$~AU. The derived orbital and stellar parameters are reported in Table~\ref{orbital_params}.

\begin{figure}
\centering
\includegraphics[width=\columnwidth]{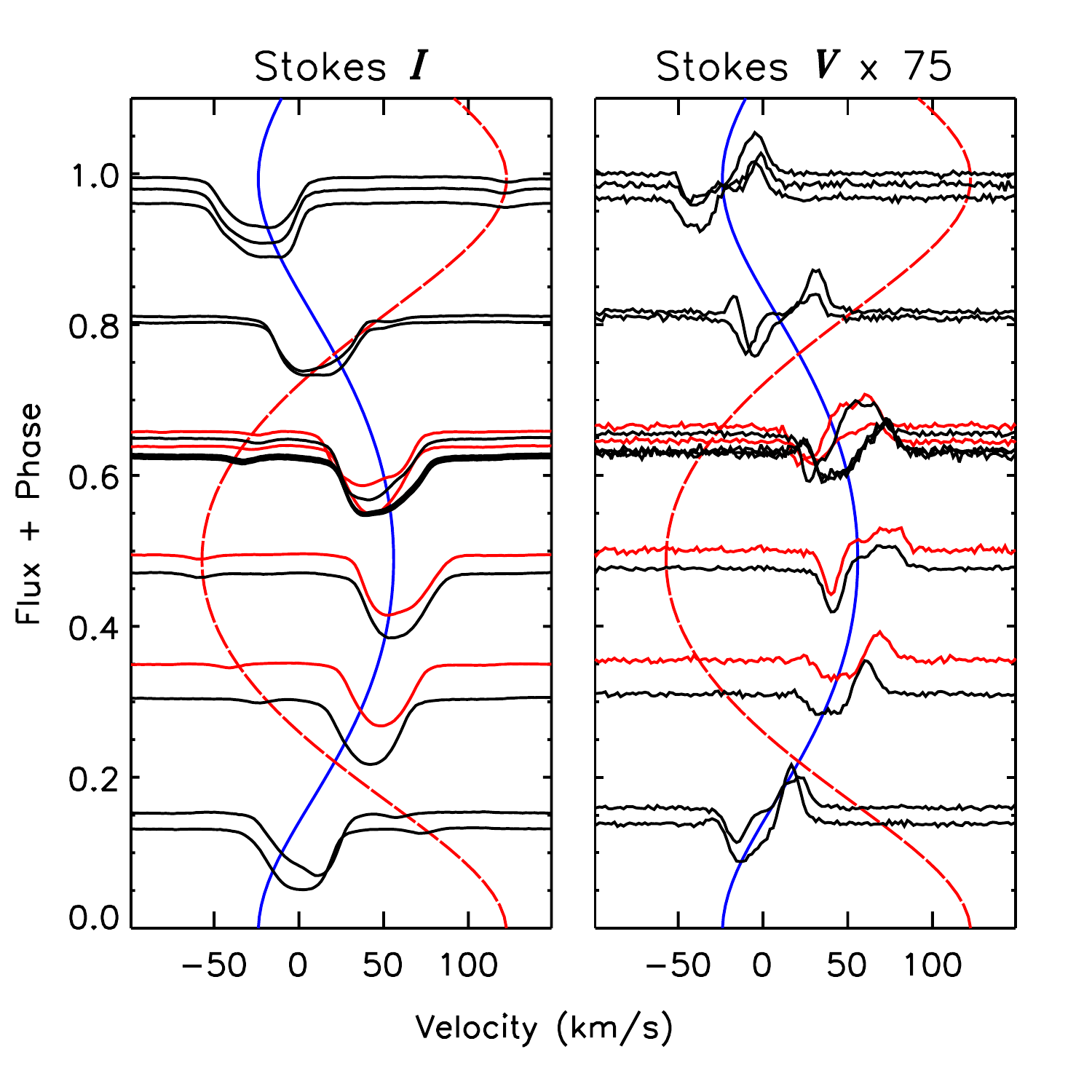}
\caption{LSD profiles phased according to the orbital ephemeris described in the text. Solid curves represent the theoretical RV variations described in Sect.~\ref{fig:Orbital Modeling}. Black line profiles were obtained in 2018. Red line profiles represent those obtained in 2013.} 
\label{fig:ProfilesOrbitPhased}
\end{figure}

Figure \ref{fig:ProfilesOrbitPhased} illustrates the LSD Stokes $I$ and $V$ profiles stacked according to orbital phase and overlaid with the predicted RV variations according to the adopted orbital model. The primary's profile is traced by the blue curve, while the subtle secondary profile is traced by the red curve. It is clear in the right-hand frame that the Stokes $V$ profiles clearly coincide with the primary's Stokes $I$ profile at all orbital phases. No Stokes $V$ profile is observed in association with the secondary's profile. In addition, it is clear that the primary's Stokes $V$ signatures show diverse shapes at similar orbital phases. This strongly suggests that the primary's rotational period is {\em not} equal to the orbital period, in contradiction to the proposals of \citet{1980ApJS...42..421B} and \citet{2017A&A...601A..14M}. 

As discussed above, the frequency of 0.16806\,d$^{-1}$ detected in the TESS data corresponds to the orbital frequency. It is therefore likely that the orbital frequency is detected in the TESS photometry. We detrended the data to remove the lowest frequency (0.033\,d$^{-1}$) and added the discussed frequencies to the model, also including $f_{\rm orb}$ and $2\,f_{\rm orb}$ (the latter barely seen in the frequency spectrum). The amplitudes are $\sim$0.87 and 0.38 mmag. We used the Wilson-Devinney lightcurve modeling code \citep{1971ApJ...166..605W} to compute the predicted lightcurve based on the derived orbital and physical parameters. Both the observed amplitudes and phases of the variability components are acceptably explained by the main contributing proximity effects from the distortion of the primary (with $2\,f_{\rm orb}$) and reflection effect for the secondary (with $f_{\rm orb}$). 

\section{Fundamental parameters}\label{sect:fund_param}

The fundamental parameters of the primary component (effective temperature $T_{\rm eff}$, luminosity $L$, radius $R$, and evolutionary mass $M$) were estimated by modelling the observed spectral energy distribution (SED) and the observed line profiles, then situating the star on the Hertzsprung-Russell diagram (HRD). As discussed in the following sections, we considered models consisting of (1) only the primary component and (2) both the primary and secondary components.

\subsection{SED fitting}\label{sect:sed}

Various photometric measurements of HD~25267 are available in the literature. This includes Johnson $U$, $B$, and $V$, \emph{Gaia} $G$, $G_b$, and $G_r$ \citep{Gaia2016,Gaia2018}, Geneva \citep{Rufener1988}, 2MASS \citep{Cohen2003}, and WISE \citep{wright2010} measurements. We used the {\sc pyphot} Python package\footnote{https://github.com/mfouesneau/pyphot} to convert each measurement from magnitudes to units of erg\,s$^{-1}$\,cm$^{-2}$\,{\AA}$^{-1}$. A distance of $d=95.6\pm1.9$\,pc was inferred from the reported Gaia parallax measurement of $10.67\pm0.21$\,mas \citep{Lindegren2018}; based on the relatively low distance, reddening produced by the interstellar medium is assumed to be negligible \citep[e.g.][]{Vergely1998}.

The observed flux measurements were fit using the Markov chain Monte Carlo (MCMC) routine {\sc emcee} \citep{foreman-mackey2013a}. Samples were drawn by linearly interpolating the grid of synthetic spectral energy distributions (SEDS) computed by \citet{coelho2014}. This grid spans a range of effective temperatures ($3000\leq T_{\rm eff}\leq25000\,{\rm K}$) and surface gravities ($-0.5\leq\log{g}\leq5.5\,{\rm (cgs)})$ and includes a range of metallicities characterized by [$\alpha$/Fe] and [Fe/H]; we used the [$\alpha$/Fe$]=0$ and $-1.0\leq{\rm [Fe/H]}\leq0.2$ models. Each interpolated model was scaled by a factor $(R/d)^2$ where $R$ corresponds to the star's radius.

The MCMC analysis was carried out using both a single-star and double-star model. In both analyses we adopted a Gaussian [Fe/H] prior defined by the ${\rm [Fe/H]}=0.0\pm0.2$ value derived for a large sample of FGK dwarfs within the solar neighbourhood \citep{casagrande2011}. The [Fe/H] values for each star in the double-star model were assumed to be equal and thus, were characterized by a single [Fe/H] value. An initial mass function \citep{chabrier2003} prior was also included by deriving the stellar masses associated with each model from $\log{g}$ and $R$. In the case of the double-star model, the primary component's $T_{\rm eff}$ and $R$ values were forced to exceed those of the secondary component (i.e. $T_{\rm eff,A}>T_{\rm eff,B}$ and $R_{\rm A}>R_{\rm B}$).

\begin{figure}
\centering
\includegraphics[width=\columnwidth]{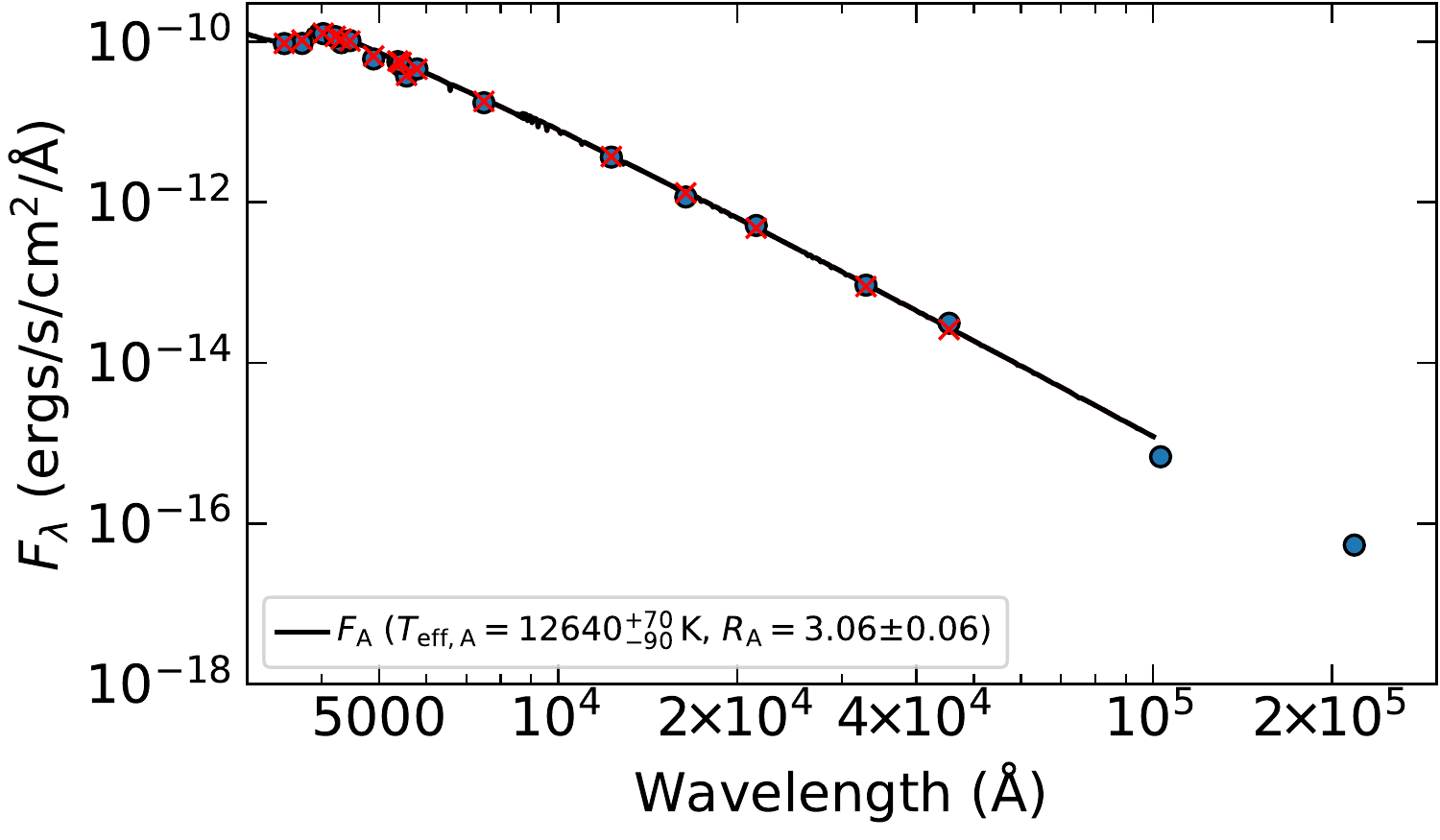}\vspace{-0.4cm}
\caption{The best-fitting single-star model compared with the photometric measurements (blue circles). The solid black curve corresponds to the total flux and the red `$\times$' symbols indicate the model flux associated with each bandpass. The W3 and W4 WISE measurements were not considered in the fit since they extend beyond the wavelength range of the adopted model SED grid, however, they are included here for completeness.}
\label{fig:sed_fit}
\end{figure}

The best-fitting parameters and their associated $1\sigma$ uncertainties were estimated from the marginalized posterior probability densities. We find that $T_{\rm eff,A}$ and $R_{\rm A}$ are consistent for both the single- and double-star models; the $T_{\rm eff,B}$ and $R_{\rm B}$ marginalized distributions yielded maximum-probability values of $11200\,{\rm K}$ and $0.6\,R_\odot$ with associated uncertainties $\sim$2000\,K and $\sim$0.5\,$R_\odot$. The statistical significance of the double-star model with respect to the single-star model was evaluated by comparing each fit's Bayesian information criterion (BIC) \citep{schwarz1978} where $\Delta{\rm BIC}\gtrsim10$ is considered significant \citep{kass1995}. We find $\Delta{\rm BIC}<2$ and therefore, the improvement of the fit yielded by the double-star model relative to the single-star model is not considered to be significant. We therefore only consider the best-fitting single-star model parameters for the primary component of $T_{\rm eff,A}=12640_{-90}^{+70}\,{\rm K}$, $R_{\rm A}=3.06\pm0.06\,R_\odot$, $\log{g}=3.4\pm0.1\,{\rm (cgs)}$, and ${\rm [Fe/H]}=0.1\pm0.1$. In Fig. \ref{fig:sed_fit}, we show the best-fitting single-star model.

\subsection{Spectral line fitting}\label{sect:Balmer_fit}

We carried out a spectral modelling analysis in order to provide additional constraints on HD~25267's fundamental parameters and verify those values derived from the SED. As with the SED modelling analysis, we consider both single-star and double-star models. We used the open-source Dynamic Nested Sampling Python package {\sc dynesty} \citep{speagle2020} to generate posterior probability distributions. Although slower than the MCMC algorithms employed by {\sc emcee}, {\sc dynesty} is well-suited to cases involving complex and multi-modal distributions.

Grids of synthetic spectra were generated using the Grid Search in Stellar Parameters ({\sc gssp}) code \citep{tkachenko2015}, which uses the LTE-based radiative transfer code {\sc SynthV} \citep{tsymbal1996} and pre-computed {\sc LLmodel} model atmospheres \citep{shulyak2004}. Two grids of synthetic spectra were generated based on the SED fitting analysis presented above: one with $9000\leq T_{\rm eff}\leq16000\,{\rm K}$ and one with $4000\leq T_{\rm eff}\leq12000\,{\rm K}$, which correspond to the primary and secondary components, respectively. Both grids spanned $3.0\leq\log{g}\leq5.0\,{\rm (cgs)}$ and $-0.8\leq{\rm [M/H]\leq+0.8}$; instrumental broadening assuming $R=65000$ was included in the models while microturbulent broadening was fixed at $0\,{\rm km\,s}^{-1}$. Based on the grid of {\sc LLmodel} atmospheres, we used grid resolutions of $\Delta T_{\rm eff}=500\,{\rm K}$, $\Delta\log{g}=0.2\,{\rm (cgs)}$, and $\Delta{\rm [M/H]}=0.2$. Models corresponding to intermediate parameters were computed by linearly interpolating the grids of synthetic spectra. In order to reduce the computation time, rotational broadening was added by convolving the models using the {\sc fastRotBroad} function of the {\sc PyAstronomy} Python package\footnote{https://github.com/sczesla/PyAstronomy}. Linear limb-darkening coefficients were estimated by interpolating the grid computed by \citet{diaz-cordoves1995}.

The contribution of the secondary component is weak throughout the majority of the observed spectra; however, absorption associated with each component's H$\alpha$ line and O\,{\sc i} triplet near 7\,770\,{\AA} is discernible by eye. We attempted to model (1) H$\alpha$ in order to constrain each component's $T_{\rm eff}$, $\log{g}$, ${\rm [M/H]}$, and ratio of the radii ($R_{\rm B}/R_{\rm A}$) and (2) the O\,{\sc i} triplet in order to constrain each component's $v\sin{i}$. (The significant non-LTE contributions affecting the O\,{\sc i} triplet make it unsuitable for more sophisticated interpretation using our LTE models.)

\begin{figure}
\centering
\includegraphics[width=\columnwidth]{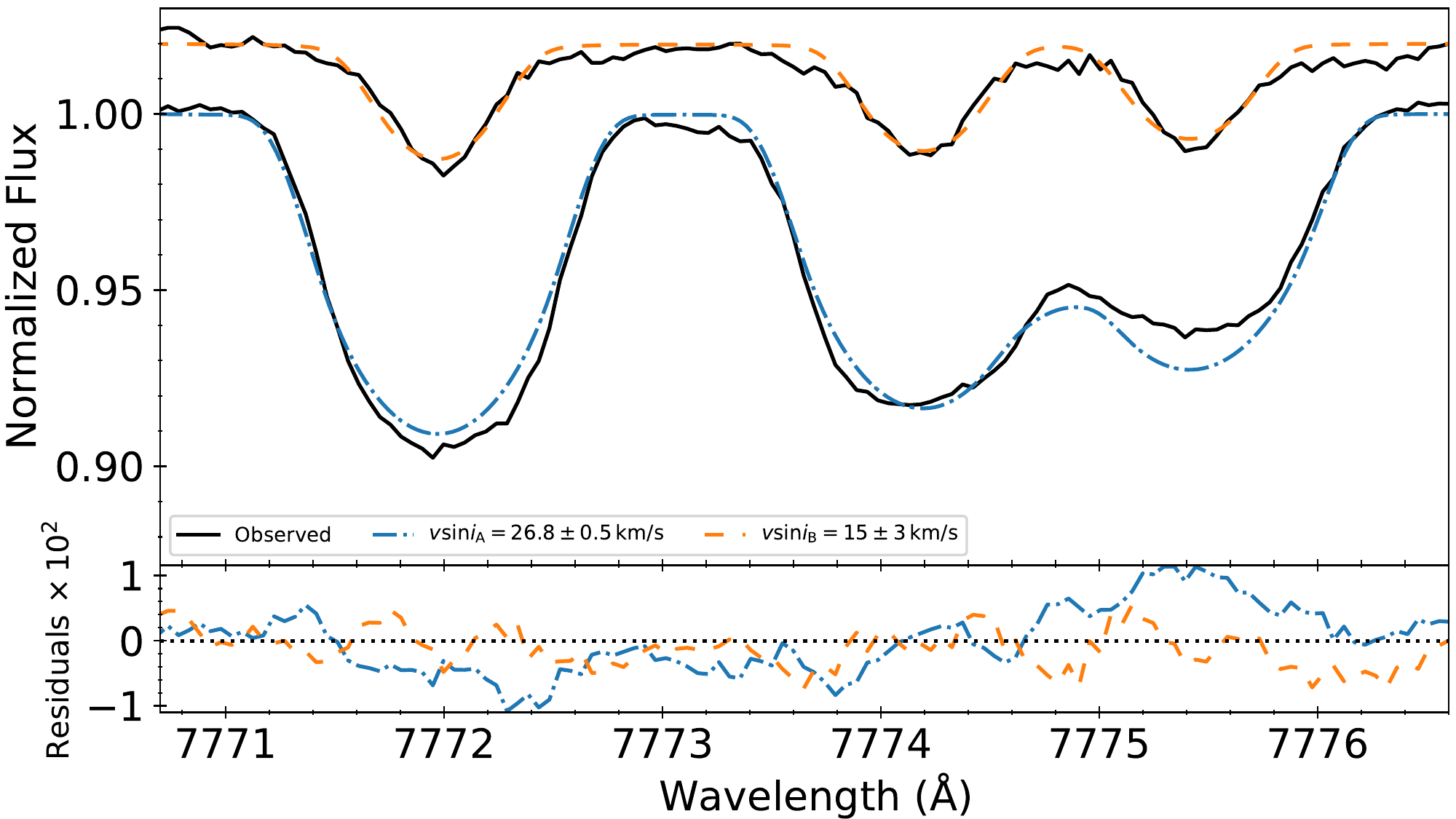}
\includegraphics[width=\columnwidth]{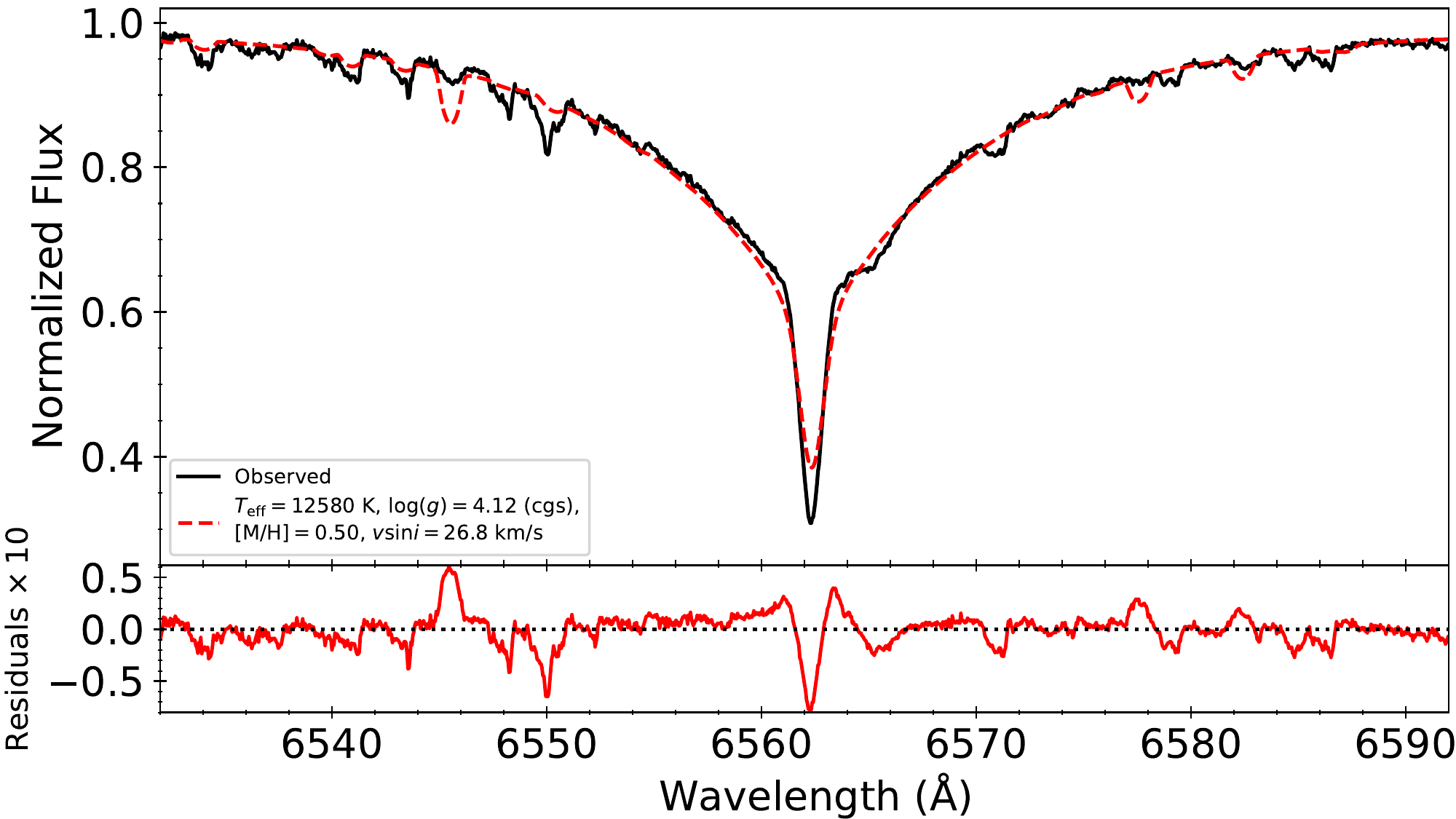}
\caption{\emph{Top:} Comparison between the disentangled O\,{\sc i} triplet and the model spectra used to derive $v\sin{i}$ of each component. The residuals are shown in the lower panel. \emph{Bottom:} Best-fitting synthetic spectrum (dashed red) compared with the observed spectrum (solid black) for H$\alpha$.}
\label{fig:spec_fit}
\end{figure}

The O\,{\sc i} triplet modelling was carried out first using disentangled spectra, which were obtained with the help of the code described by \citet{folsom:2010} and applied in several other studies \citep{kochukhov:2018,kochukhov:2019}. The fixed orbital solution found in Sect.~\ref{Sect:orbit} was adopted. The spectral disentangling is challenging for $\tau^9$~Eri due to the weakness of the spectral contribution of the secondary and the significant intrinsic variability of the primary. The O\,{\sc i} triplet is one of the few spectral features for which disentangling yields satisfactory results. Each component's $v\sin{i}$ was estimated by modelling the disentangled spectra independently. The effective temperatures were fixed at the most-probable values inferred from the double-star SED fitting ($T_{\rm eff,A}=12690$\,K and $T_{\rm eff,B}=11200$\,K) while $\log{g}$ and ${\rm [M/H]}$ were fixed at $4.0\,{\rm (cgs)}$ and $0.0$, respectively. A scaling factor was included as a free parameter to account for continuum dilution. The {\sc dynesty} code yielded $v\sin{i}_{\rm A}=26.8\pm0.5$\,{\kms} and $v\sin{i}_{\rm B}=15\pm3$\,{\kms} as estimated from the posterior probability distributions. The derivation of $v\sin{i}_{\rm B}$ was also carried out using a $T_{\rm eff}=7530$\,K based on an estimate of the secondary's temperature inferred from its HRD position (see Sect.~\ref{sect:HRD} below), which yielded the same value. The fits to the disentangled O\,{\sc i} triplet are shown in Fig.~\ref{fig:spec_fit} (top).

The modelling of H$\alpha$ was carried out using the observed spectra obtained at an epoch at which the two spectral components were found to exhibit some of the largest relative radial velocities ($\phi_{\rm orb}=0.968$). We selected a 60\,{\AA}-width region approximately centered on H$\alpha$; the spectral window was normalized by fitting a first order polynomial to the continuum located near the boundaries. Similar to the SED-fitting analysis, two models were adopted: one consisting of just the primary component and one consisting of both components. The single-star model consists of three free parameters ($T_{\rm eff,A}$, $\log{g}_{\rm A}$, and ${\rm [M/H]}$) while the double-star model includes an additional four free parameters ($T_{\rm eff,B}$, $\log{g}_{\rm B}$, ${\rm [M/H]}_{\rm B}$, and $R_{\rm B}/R_{\rm A}$); $v\sin{i}_{\rm A}$ and $v\sin{i}_{\rm B}$ were fixed at the values derived from the O\,{\sc i} modelling. Uniform priors were adopted but only $R_{\rm B}/R_{\rm A}<1$ solutions were permitted.

The single-star model yielded $T_{\rm eff,A}=12580_{-120}^{+150}\,{\rm K}$, $\log{g}_{\rm A}=4.12_{-0.04}^{+0.05}\,{\rm (cgs)}$, and ${\rm [M/H]}_{\rm A}=0.5\pm0.1$, which are consistent with the values derived using the double-star model. We note that the primary component exhibits the strong atmospheric chemical peculiarities common to Bp stars. Since the peculiar abundances are not well represented by a scaled solar abundance table, the global metallicity parameter ${\rm [M/H]}$ has little physical meaning. A more sophisticated chemical abundance analysis will eventually be required to characterize the detailed chemical composition of the primary component's atmosphere.

As with the SED fitting analysis, we evaluated the statistical significance of the double-star model using the BIC. Notwithstanding that the secondary contributes discernibly to the H$\alpha$ profile, we find $\Delta{\rm BIC}=1.2$ implying that the single-star model is preferred. This is likely an indication that the accuracy (in particular the broadening and lack of non-LTE effects) of our H$\alpha$ profile models is insufficient to achieve a formal improvement of the quality of the fit.

The single-star model fit to H$\alpha$ is shown in Fig. \ref{fig:spec_fit} (bottom).

Given that both the SED and H$\alpha$ modelling analyses did not yield a statistically significant improvement to the fits using the double-star model relative to the single-star model, we conclude that our analyses do not provide useful fundamental parameter constraints (aside from $v\sin{i}_{\rm B}$) for the secondary component.

\subsection{Hertzsprung-Russell diagram}\label{sect:HRD}

\begin{figure}
\centering
\includegraphics[width=\columnwidth]{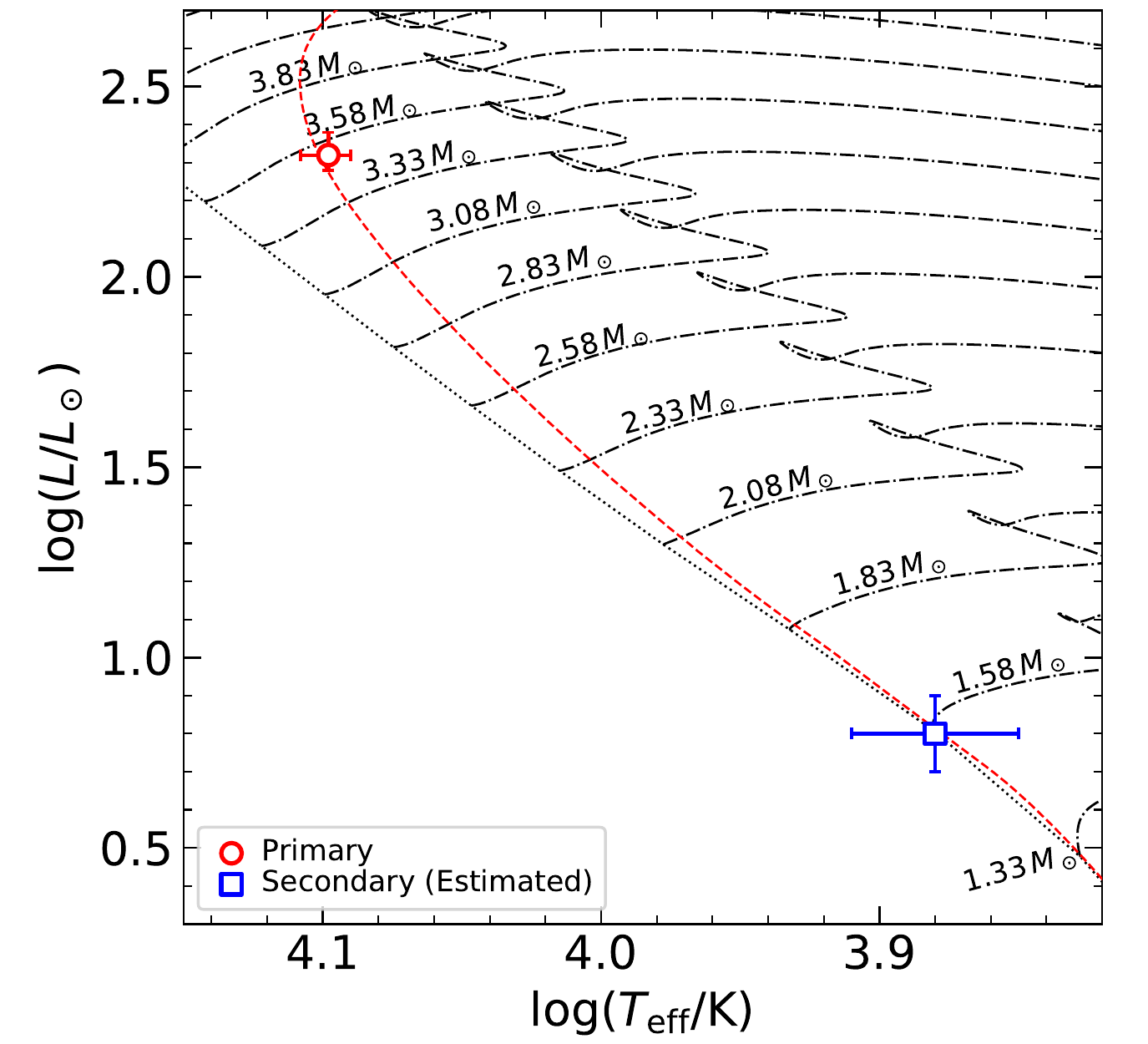}
\caption{HRD position of the primary component (red circle) and its best-fitting isochrone (dashed red). The position of the secondary component (blue square) is estimated using the evolutionary models (see Sect. \ref{sect:HRD}). The dot-dashed black lines correspond to the the solar metallicity, non-rotating MIST stellar evolutionary models computed by \citet{choi2016}; the dotted black line corresponds to the ZAMS.}
\label{fig:hrd}
\end{figure}

The mass and age of HD~25267's primary component were derived by locating the star on the HRD using {\sc emcee} \citep{foreman-mackey2013a} in conjunction with the MESA Isochrones \& Stellar Tracks (MIST) grid of evolutionary models \citep{choi2016,dotter2016}. This grid has been computed using the Modules for Experiments in Stellar Astrophysics (MESA) code \citep{paxton2011,paxton2013,paxton2015} over a wide range of masses ($0.1\leq M/M_\odot\leq300$) and Fe abundances ($-4\leq{\rm [Fe/H]}\leq0.5$). Interpolation of the model evolutionary tracks was carried out using the {\sc isochrones} Python package \citep{2015ascl.soft03010M}, which interpolates over mass, [Fe/H], and the so-called ``equivalent evolutionary point" (i.e. evolutionary points defined by specific phases such as the zero age and terminal age main sequence).

Three priors were adopted for the MCMC fitting: a Gaussian [Fe/H] prior based on the $0.0\pm0.2$ value reported for solar neighbourhood FGK dwarfs \citep{casagrande2011}, an initial mass function prior \citep{chabrier2003}, and an age prior \citep[Eqn. 17 of ][]{angus2019}. Using $T_{\rm eff,A}$ derived from the spectral modelling analysis and $R_{\rm A}$ derived from the SED modelling analysis, we derived a mass and age for the primary component of $M_{\rm A}=3.6_{-0.2}^{+0.1}\,M_\odot$ and $t_{\rm age,A}=140_{-30}^{+40}$\,Myr where the uncertainties correspond to $1\sigma$.

As noted in the previous section, no conclusive constraints for the secondary component's effective temperature and radius were able to be derived from either the SED or H$\alpha$ modelling. We attempted to roughly estimate $T_{\rm eff,B}$ and $R_{\rm B}$ based on the grid of evolutionary models. This involved using $M_{\rm A}$ derived above and $M_{\rm A}/M_{\rm B}$ derived from the radial velocity analysis presented in Sect. \ref{Sect:orbit} to obtain $M_{\rm B}=1.6\pm0.1\,M_\odot$. We adopted this $M_{\rm B}$ value and assumed an age equal to that derived for the primary component and carried out the same MCMC analysis, which yielded $T_{\rm eff,B}=7530_{-510}^{+580}$\,K and $R_{\rm B}=1.5\pm0.1\,R_\odot$. In Fig. \ref{fig:hrd}, we plot the primary component's location and the secondary component's estimated location on the HRD. These are compared with the solar metallicity, non-rotating MIST grid.

Finally, we computed a synthetic SED and spectrum of the system's H$\alpha$ region assuming the secondary properties derived above and confirmed that they are consistent with the general lack of a strong contribution of the secondary's light in the observations. 

\begin{figure*}
\centering
\includegraphics[angle=-90,width=0.325\textwidth]{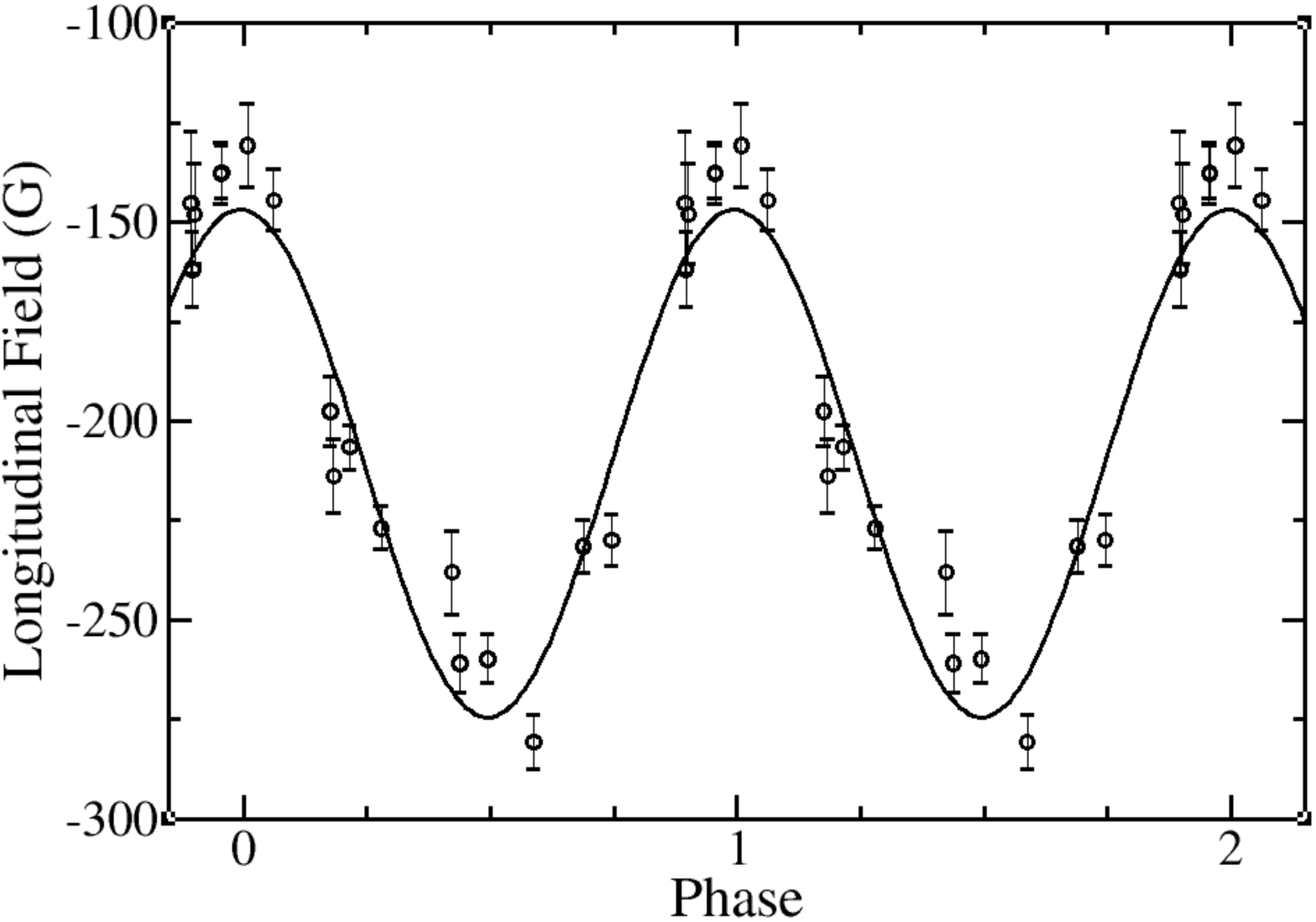}
\includegraphics[angle=-90,width=0.325\textwidth]{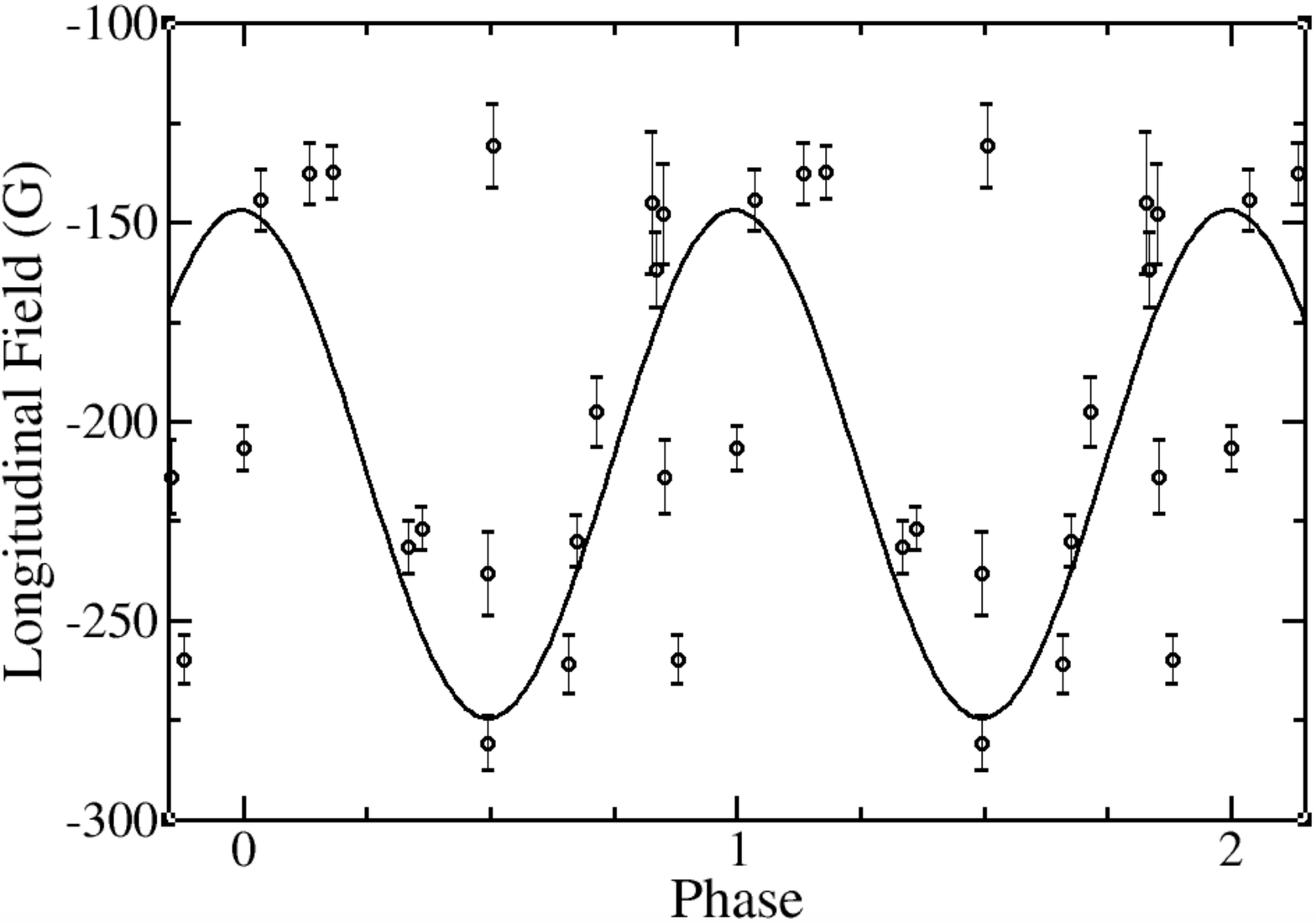}
\includegraphics[angle=-90,width=0.325\textwidth]{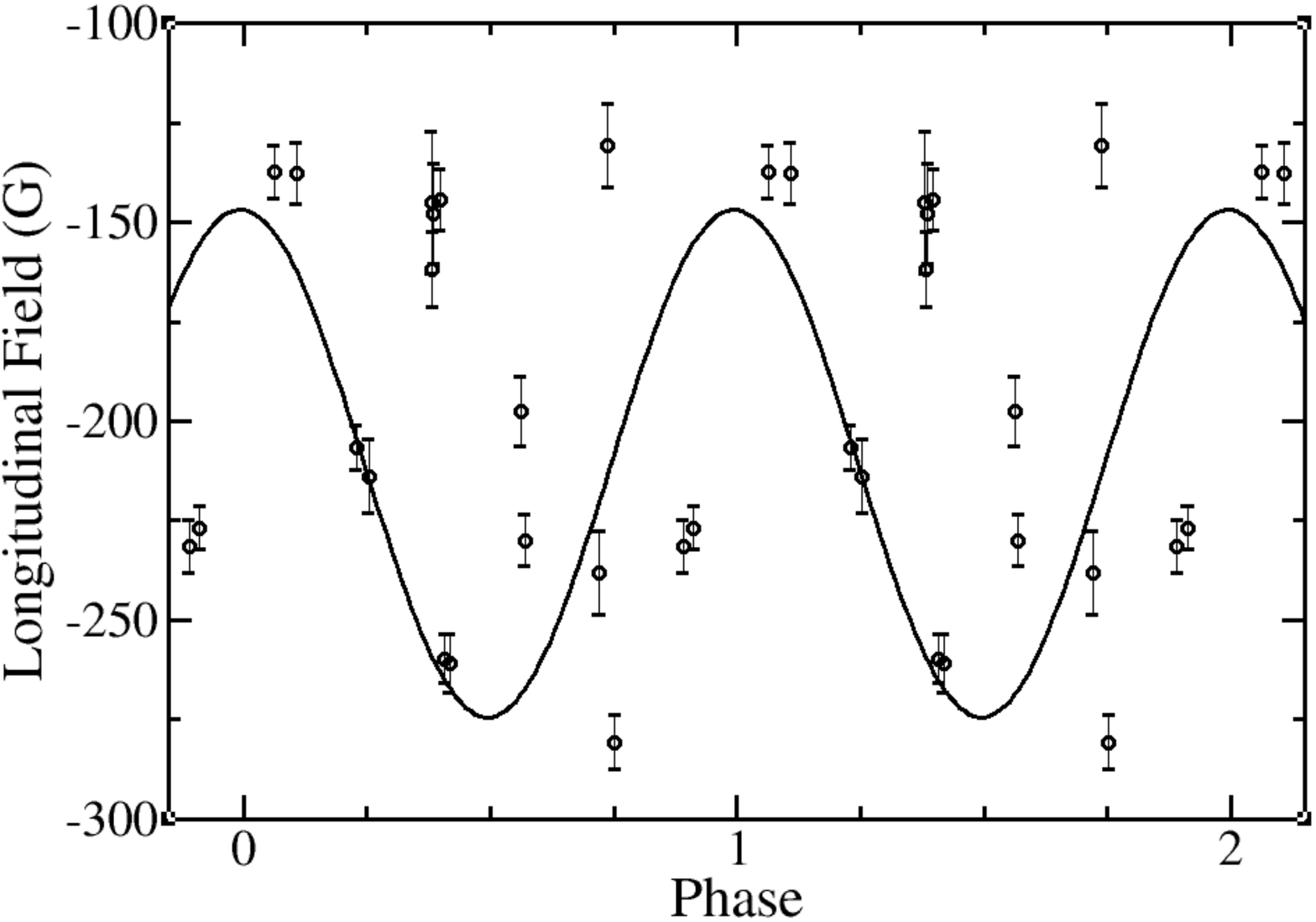}
\caption{\label{fig:BzPhasePlots}From left to right: $B_\ell$ measurements phased to rotational periods of $3.82262$~d, $1.209911$~d, and $5.95382$~d. Curves are sinusoidal fits to the phased data.}
\end{figure*}

\begin{figure}
\centering
\includegraphics[width=\columnwidth]{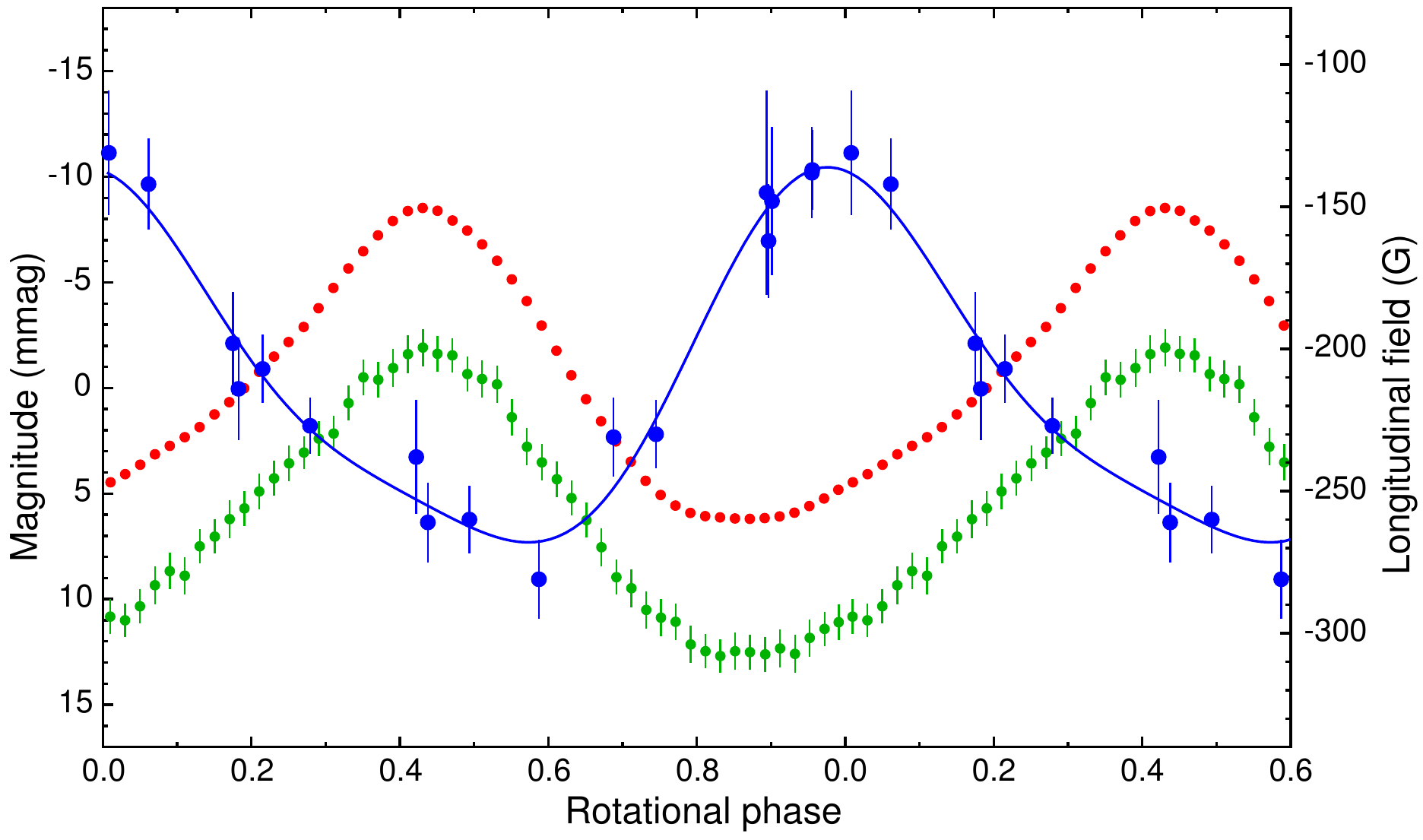}
\caption{Rotational variability of the longitudinal field and photometric brightness of $\tau^9$~Eri. All data were phased with rotational period according to Eq.~(3). Red and green curves represent TESS and SMEI photometry, respectively. The points are average values in 0.02-cycle phase intervals. Prior to phasing, the photometries were freed from the contribution of terms other than $f_{\rm rot}$ and its harmonic. Blue dots are the longitudinal magnetic field measurements. The fit consisting of two sinusoidal terms with $f_{\rm rot}$ and $2f_{\rm rot}$ is shown with the blue line.}
\label{fig:bzfitted}
\end{figure}

\begin{table}
\caption{Adopted fundamental parameters of the primary and secondary components derived in Sect. \ref{sect:fund_param}. $M_{\rm B}$ was derived by combining the primary component's evolutionary mass with the dynamical mass ratio listed in Table~\ref{orbital_params}; $T_{\rm eff,B}$, $R_{\rm B}$, and $\log{(L_{\rm B}/L_\odot)}$ are estimated from evolutionary models (see Sect. \ref{sect:HRD}). The orbital inclination, $i$, was inferred via comparison of the primary's evolutionary mass and the dynamical mass from Table~\ref{orbital_params}.
\label{tbl:fund_param}}
\centering
\begin{tabular}{lrr}
\hline
Parameter & Primary & Secondary \\
\hline
$T_{\rm eff}$ (K)   & $12580_{-120}^{+150}$ & $7530_{-510}^{+580}$ \\
$R$ ($R_\odot$)     & $3.06\pm0.06$         & $1.5\pm0.1$ \\
$\log{(L/L_\odot)}$ & $2.32\pm0.03$         & $0.8\pm0.1$ \\
$v\sin{i}$ ({\kms})          & $26.8\pm0.5$          & $15\pm3$ \\
$t_{\rm age}$ (Myr) & $140_{-30}^{+40}$     & \\
$M$ ($M_\odot$)     & $3.6_{-0.2}^{+0.1}$   & $1.6\pm0.1$ \\
$i$ ($\degr$)       & $40\pm 1$             & \\
\hline
\end{tabular}
\end{table}

\section{Rotation period and magnetic geometry of the Bp star}\label{sect:rotandmag}

The period analysis presented in Sect.~\ref{Sect:periods} and summarized in Table~\ref{tab:periods} indicates that the 5.95~d orbital period is detected in the RV data and possibly weakly in the TESS data. The $B_\ell$ and EW measurements appear to show significant modulation only according to the 3.82~d period, while the photometry shows clear signal arising from both the 3.82~d period and the 1.21~d period, with no evidence for the 5.95~d period.

To underscore these results, in Fig.~\ref{fig:BzPhasePlots} we show the $B_\ell$ measurements phased according to each of the periods mentioned above. This figure shows clearly that while the magnetic measurements phase well with the 3.82~d period, no coherent variation is achieved for either the 1.21~d or 5.95~d periods. As a consequence we conclude that the 3.82~d period corresponds to the rotation period of the magnetic Bp star primary. We therefore adopt the rotational ephemeris:
\begin{equation}\label{eq:mag}
    T_{B_\ell,{\rm max}}(E_{\rm rot}) = \mbox{HJD}\,2456528.73(5) + 3.82262(4)\cdot {E_{\rm rot}},
\end{equation}
where $E_{\rm rot}$ represents the number of rotational periods from the initial epoch at maximum $B_{\ell}$ field strength. We have adopted the precise SMEI period as the rotational period. When phased with this ephemeris, the $B_\ell$ measurements describe an approximately sinusoidal variation. The phased measurements are shown with a superimposed least-squares fit in Fig.~\ref{fig:bzfitted}, along with the extracted $3.82$~d photometric variation. The reduced $\chi^2$ of a first-order fit is 4.1, versus 1.8 for a second-order fit. This implies that non-sinusoidal variability, likely introduced by the presence of non-uniform surface chemical abundance distributions, is present in the longitudinal field curve.

Combining the adopted rotation period with the inferred stellar radius ($3.06\pm 0.06~R_\odot$; Table~\ref{tbl:fund_param}) and projected rotational velocity ($26.8\pm 0.5$~{\kms}; Sect.~\ref{sect:Balmer_fit}) implies a rotation axis inclination of $i=41\pm 2\degr$. The similarity between this value and the inferred orbital inclination will be discussed in further detail in Sect.~\ref{sect:discussion}. 

We can infer the magnetic field strength and geometry of the primary's dipole component directly from the longitudinal field variation. We modeled the phase $B_\ell$ curve using Landstreet's {\sc fldcurv} code. Assuming a limb-darkening coefficient of 0.3 and adopting $i=41\pm 2\degr$ above we derive $\beta=158\pm 5\degr$ and $B_{\rm d}=1040\pm 50$~G. Hence the primary star is inferred to have a moderately strong ($\sim$1~kG) dipole that is slightly ($\sim$20$\degr$) offset from its rotation axis.

\section{Magnetic field of the secondary star}\label{sect:secmag}

The secondary component shows no evidence of circular polarisation associated with its mean line in the 16 observations that are unblended with the primary. We measured the longitudinal field from each observation using an integration range of $\pm 27$ {\kms} about the centre-of-gravity of the profile. All are consistent with zero field, with a median $B_\ell$ error bar of about 18~G, and a best $B_\ell$ error bar of 12~G. We conclude that the secondary component of \tn\ is unlikely to have a surface dipole magnetic field stronger than about 10 times the median error bar, i.e. a few hundred gauss. However, we can arguably obtain a more realistic constraint if we assume that the secondary's rotational axis is aligned with the orbital axis. This is a reasonable assumption if the primary exhibits aligned rotation, since the lower-mass secondary is expected to achieve orbital alignment first. In this case we obtain an upper limit of about 250~G, assuming an obliquity of $\beta=90\degr$.

\section{Origin of the 1.21 day period}\label{sect:1.21dperiod}

{With the 5.95~d period established as the orbital period, and the 3.82~d period established as the primary's rotational period, we are left to explain the origin of the 1.21~d period (corresponding to $f_1$). This period, previously reported by \citet{1985A&A...144..251M}, \citet{1991A&A...248..179C}, and \citet{2020MNRAS.493.3293B}, heretofore remained a mystery. However, the temperature of the primary component, coupled with the compatibility of this period with the known range of $g$-mode frequencies and the detection of additional frequencies in this range (since $g$-mode pulsation is frequently detected in multiple independent modes) in two independent data sets, strongly suggests that the 1.21~d period corresponds to a $g$-mode. $g$ modes originate in mid B-type stars (SPB stars) as well as F-type stars ($\gamma$ Dor stars). However, the detection of the difference frequency $f_6$ firmly establishes the association of this frequency with the magnetic Bp star. This conclusion is also supported by the large intensity of the 1.21~d modulation, given the rather small contribution of the secondary star to the system's flux.}

{For completeness we point out that the 1.21~d period could also potentially be associated with the rotation of the secondary star. However, adopting this period as its rotational period, the measured $v\sin i$ and inferred radius of the secondary (15~{\kms} and $1.5~R_\odot$ respectively; Table~\ref{tbl:fund_param}) imply a very low rotation axis inclination ($\sim$10$\degr$) which is geometrically unlikely and difficult to reconcile with the large observed photometric amplitude.}

We therefore conclude that the primary component of HD\,25267 is a magnetic SPB star, and the dominant 1.21~d modulation is the most intense of several $g$-modes that are detected in its photometric amplitude spectrum.

The residual variability in the TESS photometry is likely due to $g$-modes as the density of the spectrum of $g$-modes is very high. Whether some of this variability is also contributed by $g$-modes in the secondary is an open question and cannot be ruled out.

\section{Summary, discussion and conclusions}\label{sect:discussion}

We have analyzed 17 ESPaDOnS spectropolarimetric observations of the bright, nearby spectroscopic binary \tn, supported by Hipparcos, SMEI and TESS photometry. We have discovered the weak mean spectral line of the secondary component, and detect a strong magnetic field in the primary component. We conclude that \tn\ consists of a late-type main sequence Bp star ($M = 3.6_{-0.2}^{+0.1}~M_\odot$, $T_{\rm eff} = 12580_{-120}^{+150}$~K) with a rotation period of $3.82262(4)$ days in a mildly eccentric $5.95382(2)$-day orbit with an late A/early F main sequence secondary companion ($M = 1.6\pm 0.1~M_\odot$, $T_{\rm eff} = 7530_{-510}^{+580}$~K). The Bp star's magnetic field is approximately dipolar with Oblique Rotator parameters $i=41\pm 2\degr$, $\beta=158\pm 5\degr$ and $B_{\rm d}=1040\pm 50$~G. 

Our detection of the secondary's spectral lines establishes \tn\ as the nearest and brightest spectroscopic binary containing a magnetic Ap/Bp star.

One interesting feature of this system is the similarity between the inclination of primary's rotation axis and the orbital axis, consistent with spin-orbit alignment. Since the lower-mass secondary star likely experienced spin-orbit alignment before the primary, its rotation axis inclination is also likely $\sim$40$\degr$. In this case its derived radius and $v\sin i$ indicate that it has a rotation period similar to that of the primary. Given that the system has clearly not achieved spin-orbit synchronization, we are unable to provide a physical reason for the similarity of the measured primary rotation period and inferred secondary rotation period, and conclude that it must be a coincidence. 

A second interesting feature of this system is the near anti-alignment of the primary's magnetic and rotation axes. \citet{2019MNRAS.488...64P} studied the strength of tidal and electromagnetic interactions in the doubly-magnetized binary system  $\epsilon$~Lup. They derived that in that system, the lowest-energy configuration for the magnetic axes has them anti-aligned and parallel with the rotation axes. \tn\ is in many ways very similar to the system HD\,98088 \citep{2013MNRAS.431.1513F}. Both have orbital periods near 6~d, both are moderately eccentric, and both contain a magnetic Ap/Bp primary and a less massive and less luminous secondary. \citet{2013MNRAS.431.1513F} concluded that the magnetic axis of the Ap star HD\,98088A is nearly perpendicular to its rotation axis, and that its magnetic axis points approximately at the secondary star as the two stars orbit. (This is only possible because HD\,98088 exhibits synchronized orbital and rotational periods.) Hence HD\,25267A provides another rare and interesting datum concerning the magnetic geometries of magnetic stars in close binary systems.

{An additional period of $1.21$~d that is clearly detected in all photometric datasets is interpreted as one of several likely $g$-modes detected in the SMEI and TESS lightcurves. This makes \tn\ a very interesting system in the modern context of stellar magnetism: a close spectroscopic binary containing a (multi-periodic) pulsating, magnetic star. Moreover, the detection of the 5.95~d orbital period in the TESS photometry (corresponding to $f_{10}$) suggests that the stars are sufficiently close to interact tidally. Given its brightness and proximity, this system is ripe for detailed follow-up.}

\section*{Acknowledgments}
We thank Dr.~Alexandre David-Uraz for assistance with the TESS data. We also thank Dr.~Coralie Neiner for helpful suggestions. GAW is supported by a Discovery Grant from the Natural Sciences and Engineering Research Council (NSERC) of Canada. OK acknowledges funding from the Swedish Research Council and the Swedish National Space Agency. APi acknowledges support from the National Science Centre (NCN) grant 2016/21/B/ST9/01126. This work has made use of the VALD database, operated at Uppsala University, the Institute of Astronomy RAS in Moscow, and the University of Vienna.

\section*{Data Availability}

The spectropolarimetric data underlying this article are available from the Polarbase database (http://polarbase.irap.omp.eu), and are uniquely identified with the spectrum IDs listed in Table \ref{tab:Log}. The photometric data are available from their respective public databases.

\bibliography{main} 

\bsp	
\label{lastpage}
\end{document}